\documentclass[structabstract]{aa}

\usepackage{txfonts}
\usepackage{graphicx}
\usepackage{natbib}
\bibpunct{(}{)}{;}{a}{}{,}

\begin{document}

  \title{A close look at the Centaurus A group of galaxies: \\ 
I. Metallicity distribution functions and population gradients in early-type dwarfs}
  \author{D. Crnojevi\'{c}\inst{1}\fnmsep\thanks{Corresponding \email{denija@ari.uni-heidelberg.de}. \newline Member of IMPRS (International Max Planck Research School) for Astronomy \& Cosmic Physics at the University of Heidelberg and of the Heidelberg Graduate School for Fundamental Physics.} \and E. K. Grebel\inst{1} \and A. Koch\inst{2}
}

 \institute{Astronomisches Rechen-Institut, Zentrum f\"{u}r Astronomie der
   Universit\"{a}t Heidelberg, M\"{o}nchhofstrasse 12-14, 69120 Heidelberg, Germany
 \and
Department of Physics and Astronomy, University of Leicester, University Road, Leicester LE1 7RH}

  \date{Received 9 October 2009 / Accepted 26 January 2010}

  \abstract
  {}{We study dwarf galaxies in the Centaurus A group to investigate their metallicity and possible environmental effects. The Centaurus A group (at $\sim4$ Mpc from the Milky Way) contains about 50 known dwarf companions of different morphologies and stellar contents, thus making it a very interesting target to study how these galaxies evolve.}{Here we present results for the early--type dwarf galaxy population in this group. We use archival HST/ACS data to study the resolved stellar content of 6 galaxies, together with isochrones from the Dartmouth stellar evolutionary models.}{We derive photometric metallicity distribution functions of stars on the upper red giant branch via isochrone interpolation. The 6 galaxies are moderately metal--poor ($<$[Fe/H]$>=-1.56$ to $-1.08$), and metallicity spreads are observed (internal dispersions of $\sigma$[Fe/H]$=0.10$--$0.41$ dex). We also investigate the possible presence of intermediate--age stars, and discuss how these affect our results. The dwarfs exhibit flat to weak radial metallicity gradients. For the two most luminous, most metal--rich galaxies, we find statistically significant evidence for at least two stellar subpopulations: the more metal--rich stars are found in the center of the galaxies, while the metal--poor ones are more broadly distributed within the galaxies.}{We find no clear trend of the derived physical properties as a function of (present-day) galaxy position in the group, which may be due to the small sample we investigate. We compare our results to the early--type dwarf population of the Local Group, and find no outstanding differences, despite the fact that the Centaurus A group is a denser environment that is possibly in a more advanced dynamical stage.}

\titlerunning{A close look at the Centaurus A group of galaxies: I. Early-type dwarfs}
  \maketitle


\section{Introduction}

Dwarf galaxies are fundamental ingredients for the assembly of luminous structures in our Universe. They may be the smallest baryonic counterparts of dark matter subhalo building blocks in a $\Lambda$CDM cosmology, and appear to be the most dark matter dominated objects in the Universe \citep[see e.g.][]{gilmore07}. Although their role is still under debate, it is certain that they are the most numerous type of galaxies, and an increasing number of them are being newly discovered with the current state-of-the-art surveys \citep[see e.g.][]{belokurov06, zucker06a, zucker06b, zucker07}. It is thus interesting to study their physical properties in order to gain a deeper understanding of how galaxies evolve, and of the main processes that drive this evolution.

Dwarf galaxies in groups have been looked at in detail in the last decade \citep[e.g.,][]{trentham02, kara04, kara05, grebel07, sharina08, weisz08, bouchard08, dalcanton09, koleva09}. The main purpose of most of these studies was to catalogue and characterize new objects in nearby groups and to look for possible environmental effects on galaxy evolution (see \citealt{grebel00} for details). The largest amount of information is, of course, coming from our own Local Group, for which very deep observations permit us to have a broad and detailed view of the physical properties of dwarf galaxies, and to investigate how they form and evolve in this type of environment \citep{grebel97, mateo98, vandenb99, tolstoy09}. 

In Local Group dwarf galaxies, all of the morphological types contain old populations, even though their fractions differ (e.g. \citealt{grebel97}; \citealt{mateo98}; \citealt{grebel03}; \citealt{tolstoy09}). The old populations ($\gtrsim 10$ Gyr) in the classical dwarf spheroidal companions of the Milky Way seem to share a common age within the accuracy of photometric age-dating techniques \citep{grebel04}. The main difference between early--type (spheroidal and elliptical) dwarfs and late--type (irregular) dwarfs is that only the latter contain large amounts of neutral gas and formed stars throughout their whole history. Moreover, dwarf galaxies both in the Local Group and in other groups follow the general morphology--density relation that holds in dense environments (e.g. \citealt{einasto74}; \citealt{dressler80}; \citealt{kara02}). This means that early--type dwarfs are normally found within $\sim300$ kpc from the center of the dominant galaxies in the group, while late--type dwarfs are more widely distributed.

The chemical composition of stars in early--type dwarfs (i.e., faint dwarf spheroidals and dwarf ellipticals, the latter showing higher central surface brightnesses and more elongated shapes) has been well studied with detailed spectroscopic and kinematic data. All these dwarfs are metal--poor and show wide metallicity spreads \citep[e.g.,][]{shetrone01, sara02, tolstoy04, battaglia06, helmi06, koch06, bosler07, koch07a, koch07b, gullieu09}. Their metallicity distribution functions (MDFs) often show a slow increase toward higher metallicities and then a steeper decline \citep[see][]{koch07a}, but there are many individual differences reflecting a wide range of complex star formation histories (e.g., \citealt{grebel97}; \citealt{tolstoy09}). Occasionally there is evidence for the presence of intermediate--age and even younger populations (e.g., in the dwarf spheroidals Fornax and Carina). Detailed models of chemical evolution have been developed and applied to dwarf spheroidals of our Local Group (e.g., \citealt{lanfranchi04}; \citealt{marcolini06}; \citealt{marcolini08}; \citealt{revaz09}), and they are able to reproduce the shape of the observed MDFs. The above mentioned asymmetry of the MDFs with a steeper fall-off on the metal--rich side may be explained by an evolution that is regulated by supernova explosions and stellar winds (e.g., \citealt{dekel86}; \citealt{lanfranchi04}).

Previous studies have also looked at possibly distinct stellar spatial distributions of different stellar populations in early--type dwarf galaxies \citep[e.g.,][]{stetson98, hurleyk99}. The first systematic study of a large sample of dwarfs was carried out by \citet{harbeck01}, who investigated the presence of morphological gradients based on horizontal branch and RGB stars for a sample of 9 galaxies. They showed that if gradients are present they are always such that the more metal--rich and/or younger populations are more centrally concentrated. Later on, spectroscopic studies were able to also confirm chemically, and in some cases also kinematically, distinct subpopulations for various dwarfs, like Fornax, Sculptor and Sextans (e.g., \citealt{tolstoy04}). However, there are also cases for which such distinct populations were not found, but only weak metallicity gradients as a function of radius were present \citep[e.g.,][]{harbeck01, koch06, koch07a, koch07b}. 

It is then interesting to extend our knowledge to dwarf galaxies in other groups, in order to be able to draw general conclusions about dwarf galaxy properties in different environments. Do loose, filamentary structures (like the Canes Venatici cloud or the Sculptor group, see \citealt{kara03_can, kara03_sc}) have similar properties in their galaxy populations as the denser, more evolved groups we know (the Local Group itself, or the Centaurus A group, e.g., \citealt{kara02}), or are there striking differences in the way they spend their lives? Future studies should be able to answer this and many further questions by looking closely at nearby groups, and comparing them to what we already know about our own.

The currently available telescopes and instruments do not only permit us to substantially improve the census of galaxies in nearby groups, but also to derive their detailed photometric properties, providing new insights into a range of physical properties for these objects. In particular, the deep high resolution images from the Hubble Space Telescope (HST) are now providing optically resolved stellar populations in dwarfs within $\sim10$ Mpc from the Local Group. Even though the limiting absolute magnitude dramatically decreases with their distance, we still can observe these galaxies down to comparable sensitivity levels as we used to see much closer dwarf galaxies in our own Local Group, as recently as only a decade ago \citep[e.g.,][]{dohm97}. 

The Centaurus A group is located in the southern hemisphere, at an average Galactic latitude of $b\sim20^{\circ}$. Despite this low latitude, most of its members are not highly contaminated by Galactic foreground extinction \citep[see, e.g.,][]{schlegel98}, making it a very popular target of research. It is one of the groups that are closest to our own, with a mean Galactocentric distance of $\sim3.8$ Mpc \citep{hui93, kara02, kara07, harrisg09} and with a variety of morphological types among its member galaxies. The group is dominated by the giant radio-loud elliptical galaxy Centaurus A (NGC5128), which shows a very perturbed morphology and which has probably undergone several mergers in its recent past \citep[e.g.,][]{meier89, mirabel99, kara05}. Within the Centaurus A group, there is a subgroup centered on the spiral galaxy M83 (NGC5236), even though it is not clear whether this subgroup is approaching or receding from Centaurus A \citep[e.g.][]{kara07}. The search and study of dwarf members in this group have been pursued at different wavelenghts for more than a decade \citep{cote97, kara98, banks99, cote00, jerjen00b, jerjen00a, kara02, kara04, kara05, rejkuba06, bouchard07, grossi07, kara07, lee07, bouchard08, cote09}. Now there are more than 50 confirmed dwarf galaxies known in the Centaurus A group, of which about $3/5$ are early--type dwarfs. When looking at the luminosity function of the entire group, we note that there are more luminous early--type dwarfs than in our own Local Group or in the Sculptor group, as expected for a more evolved group \citep{jerjen00b}. However, many more faint early-type dwarfs could possibly be found if more sensitive surveys were available. In particular, we know now that the Local Group contains about 20 dwarfs with $M_B>-10$, while only 4 such faint objects have been detected so far in the Centaurus A group, due to its distance.

Some of the galaxies presented here have already been investigated in previous studies (e.g. \citealt{cote97}; \citealt{jerjen00b}; \citealt{bouchard07}; \citealt{bouchard08}; \citealt{cote09}), in some cases even with the same dataset considered in our work (\citealt{kara07}; \citealt{makarova08}; \citealt{sharina08}). However, most of these studies concentrated on large samples of objects, not investigating their physical properties individually. \citet{bouchard07} find that in the Centaurus A group there is an apparent gap in HI masses: the detected galaxies have gas masses of about $10^{7}$M$_{\odot}$, or they are not detected at all (which permits them to put an upper limit of $\sim10^{6}$M$_{\odot}$). \citet{bouchard07} conclude that the Centaurus A group environment must favour an efficient gas stripping from its dwarf companions. However, they also point out that due to the limits of the HI survey, it is currently not possible to detect so called mixed--type dwarfs in the Centaurus A group (with no ongoing star formation but presence of neutral gas). Thus, knowledge of only the stellar content of early--type dwarfs does not tell us with certainty whether they are ``contaminated'' by a residual presence of gas. \citet{bouchard08} further collected literature data for dwarfs in the Centaurus A and Sculptor groups and analysed them together with new observations. Again, there is no significant presence of ongoing star formation for early--type galaxies. \citet{bouchard08} investigate the dependence of several physical properties (optical luminosity, neutral gas and H$\alpha$ content) on environment. Similarly to the Local Group, they are able to confirm that: galaxies in denser regions of these groups have, in general, lower values of HI; the star formation in these objects is lower; and they probably formed their stellar content earlier, with respect to galaxies in low density regions. However, these correlations with environment do not rule out the simultaneous impact of internal processes that act to shape their evolution. The papers that used the same dataset as in our study were mostly considering distances \citep{kara07} or scaling relations \citep{sharina08} for an extensive sample of, respectively, dwarfs in the Centaurus A group, and dwarfs in nearby groups and in the field. In our work we want to look more in detail at the metallicity and population gradients of early--type dwarfs in the Centaurus A group.

This paper is the first in a series in which we concentrate on the resolved stellar populations of dwarf galaxies in the Centaurus A group, seen through the Hubble Space Telescope. We derive their physical properties (metallicities, star formation rates, stellar spatial distributions) and try to find links with the environment they reside in, in order to put constraints on their evolution. While this paper focusses on the early-type dwarfs, we will return to the late type dwarf irregular dwarfs in two forthcoming papers (Crnojevi\'{c}, Grebel \& Cole, in prep.).

The paper is organized as follows: we describe the data in \S \ref{data}, and we present the derived results in \S \ref{cmd_sec} (color-magnitude diagrams), \S \ref{mdf_sec} and \S \ref{serrors} (metallicity distribution functions and discussion of their uncertainties, respectively), \S \ref{ssd_sec} (metallicity and population gradients). The discussion is then carried out in \S \ref{discuss}, and our conclusions are drawn in \S \ref{conclus}.


\section{Data and photometry} \label{data}

\begin{table*}
 \centering
\caption{Fundamental properties of the studied sample of galaxies}
\label{infogen}
\begin{tabular}{lccccccccc}
\hline
\hline
Galaxy&RA&DEC&$T$&$I_{TRGB}$&$D$&$A_{I}$&$M_{B}$&$M_{V}$&$\Theta$\\
&(J2000)&(J2000)&&&(Mpc)&&&\\
\hline
\object{KK189}&$13\,12\,45.0$&$-41\,49\,55$&$-3$&$24.40\pm0.07$&$4.42\pm0.33$&$0.22$&$-10.52$&$-11.99$&$2.0$\\
\object{ESO269-66, KK190}&$13\,13\,09.2$&$-44\,53\,24$&$-5$&$24.04\pm0.04$&$3.82\pm0.26$&$0.18$&$-13.85$&$-13.89$&$1.7$\\
\object{KK197}&$13\,22\,01.8$&$-42\,32\,08$&$-3$&$24.19\pm0.04$&$3.87\pm0.27$&$0.30$&$-12.76$&$-13.04$&$3.0$\\
\object{KKs55}&$13\,22\,12.4$&$-42\,43\,51$&$-3$&$24.21\pm0.03$&$3.94\pm0.27$&$0.28$&$-9.91$&$-11.17$&$3.1$\\
\object{KKs57}&$13\,41\,38.1$&$-42\,34\,55$&$-3$&$24.10\pm0.05$&$3.93\pm0.28$&$0.18$&$-10.07$&$-10.73$&$1.8$\\
\object{CenN}&$13\,48\,09.2$&$-47\,33\,54$&$-3$&$24.10\pm0.04$&$3.77\pm0.26$&$0.27$&$-10.89$&$-11.15$&$0.9$\\
\hline
\end{tabular}
\begin{list}{}{}
\item[Notes.] Units of right ascension are hours, minutes and seconds, and units of declination are degrees, arcminutes and arcseconds. The references for the reported values are \citet{kara05} and \citet{kara07}, and \citet{georgiev08} for the absolute $V$ magnitude.
\end{list}
\end{table*}

We use archival data obtained with the Wide Field Channel (WFC) of the Advanced Camera for Surveys (ACS) aboard the Hubble Space Telescope (HST). Observations with this instrument are available for 21 dwarfs in the CenA group (programmes GO-9771 and GO-10235). Each galaxy has a 1200 seconds exposure in the $F606W$ filter and a 900 seconds exposure in the $F814W$ filter. We do not consider observations carried out with the Wide Field Planetary Camera 2 (WFPC2), since its field of view is smaller, since three of its chips have lower resolution, and since it is less sensitive than the ACS. Earlier studies of dwarfs in the Centaurus A group based on WFPC2 data were published by, e.g., \citet{kara02, kara03, rejkuba06}. In this paper we present the results for 6 early--type dwarf galaxy companions of the peculiar elliptical Centaurus A. A further sample of 11 late--type dwarfs (6 companions of Centaurus A and 5 of the giant spiral galaxy M83, respectively) will be studied in two forthcoming papers of this series.

We performed stellar photometry using the ACS module of the DOLPHOT package \citep{dolphin02}, as described in detail in, for instance, \citet{dalcanton09}. The parameters chosen for the photometry follow the prescriptions of the DOLPHOT User's Guide\footnote{http://purcell.as.arizona.edu/dolphot/.}. The stars that we then want to retain from the original photometry have to simultaneously satisfy the required cuts for different quality parameters in both wavelengths. These limits are the following: each object must be classified as a good star (meaning a not too extended or too sharp object, quantified by an ``object type'' that has to be $\leq2$) and has to be recovered extremely well in the image (each star has an assigned ``error flag'' for that, which we require to be 0 in order to avoid saturation or photometry extending out of the chip), its signal to noise ratio has to be at least 5, the sharpness parameter has to be $\vert sharp \vert \le0.3$, the crowding parameter is set to a value lower than 0.5 (in mag, which quantifies how much brighter the star would be if isolated when measured), and finally the $\chi$ has to be $\le2.5$. This selection choice leaves us with fairly clean color-magnitude diagrams. For the following study, we use the $V$ and $I$-bands, converted by the DOLPHOT program from the instrumental magnitudes with the prescriptions of \citet{sirianni05}.

Extensive artificial star tests were then run with the same package to estimate the photometric uncertainties and to assess the effects of incompleteness effects. For each galaxy, we add to the images $\sim5$ times the number of observed stars (after quality cuts). The artificial stars are added and measured one at a time by the DOLPHOT routine, in order to avoid artificial crowding. They are distributed evenly across the field of view of ACS, and have a magnitude range that goes from the brightest observed stars' magnitude to $\sim1$ mag fainter than the faintest ones (to account for fainter stars possibly upscattered in the color-magnitude diagram by noise). The color range is the same as for the observed stars. After running again the photometry with the artificial stars, we apply the same quality cuts as before. We then derive completeness curves both as a function of magnitude as well as a function of radius, in order to correct star number counts (see Sect. \ref{ssd_sec}). The limiting magnitude, taken at a $50\%$ completeness level, is at $\sim27.3$ ($\sim26.8$) mag in the $V$-band for the least (most) crowded objects of our sample, and at $\sim26.4$ ($\sim26.0$) mag in the $I$-band. The mentioned galaxies are KKS55 and ESO269-66, respectively, (as can be seen from the density maps reported in Sect. \ref{ssd_sec}), with peak stellar densities of $\sim39$ and $\sim124$ stars per $0.01$ kpc$^{2}$. At a completeness level of $50\%$, the $1 \sigma$ extremes of the photometric errors amount to $\leq0.23$ mag ($\leq0.16$ mag) in $I$-band magnitude and $\leq0.30$ mag ($\leq0.21$ mag) in color, for these two galaxies. The mean photometric errors ($\pm1 \sigma$) for each galaxy are indicated by representative errorbars in the color-magnitude diagrams shown in Sect. \ref{cmd_sec}.

The main properties of the galaxies studied here are listed in Table \ref{infogen} as follows: column (1): name of the galaxy, (2-3): equatorial coordinates (J2000), (4): morphological type, (5): $I$-band magnitude at the tip of the red giant branch (from \citealt{kara07}), (6): distance of the galaxy derived by \citet{kara07} with the tip of the red giant branch method, (7): foreground extinction in the $I$-band from \citet{schlegel98}, (8): absolute $B$ magnitude from \citet{kara05}, (9): absolute $V$ magnitude from \citet{georgiev08}, and (10): tidal index, taken from \citet{kara07}.


\section{Color Magnitude Diagrams} \label{cmd_sec}

\begin{figure*}
 \centering
  \includegraphics[width=13cm]{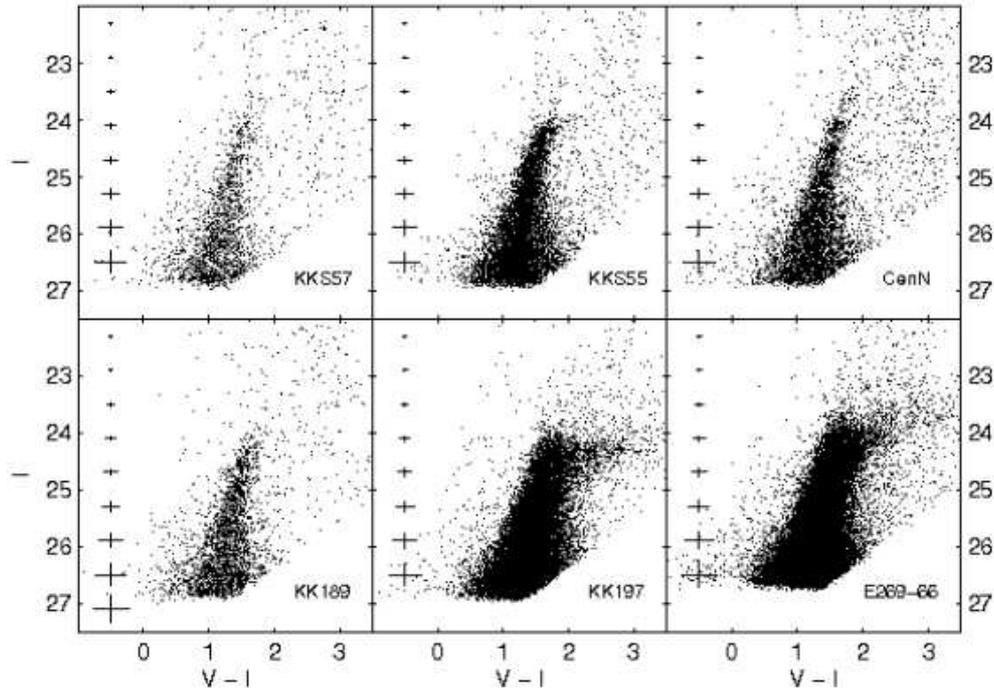}
 \caption{\footnotesize{Color magnitude diagrams of the six early--type dwarf galaxies studied in this paper, ordered by absolute $V$ magnitude of the galaxies. The main feature visible in all of the CMDs is a prominent RGB, while luminous AGB stars are less numerous (see text for an estimate for each galaxy). Representative errorbars derived from artificial star experiments are shown on the left side of the diagrams. At an $I$-band magnitude of $25.5$, the $1 \sigma$ photometric errors are of approximately $\sim0.1$ mag in magnitude and $\sim0.15$ mag in color (see text for details).}}
 \label{cmds}
\end{figure*}

The color--magnitude diagrams (CMDs) of the six galaxies are shown in Fig. \ref{cmds}, ordered by increasing absolute $V$ magnitude of the galaxies. The CMDs all show prominent red giant branches (RGBs). RGB stars may encompass stars with a wide range of ages, starting with stars as young as $1-2$ Gyr. Populations with ages of a few Gyr (``intermediate--age stars'', $1-9$ Gyr) can be recognized by a number of additional features including more luminous main sequence turnoffs, red clump and vertical red clump stars, and luminous asymptotic giant branch (AGB) stars with luminosities greater than the tip of the RGB (TRGB). Unfortunately, the HST data are not sufficiently deep to reach the main sequence turnoffs or the red clump stars of the intermediate--age populations, as these are below our detection limit of $\sim27$ $I$-band apparent magnitude (the red clump would be expected to have a magnitude of $\sim28$ in the same band). However, we can infer the presence or absence of intermediate--age populations from the presence or absence of luminous AGB stars. Even younger stars ($<1$ Gyr) are definitely not present in these galaxies, as there are no objects found in the region blueward of the RGB (i.e., upper main sequence, or massive blue and red He-burning stars). A quantitative evaluation of the amount of intermediate--age stars, and how it affects our results, will be presented in Sect. \ref{ssd_sec}. Similar evidence for intermediate--age components has also been found in some of the dwarf spheroidals and dwarf ellipticals of the Local Group, like for instance Leo I, Leo II, Fornax, Carina, NGC147 and NGC185 (e.g., \citealt{han97}).

In the CMDs of Fig. \ref{cmds}, we also show representative photometric errors derived from artificial star tests. The ($1 \sigma$) error is $\sim0.1$ mag in magnitude and $\sim0.15$ mag in color at an $I$-band magnitude of, respectively: $25.60$ for KKS57, $25.55$ for KKS55, $25.50$ for CenN, $25.55$ for KK189, $25.45$ for KK197 and $25.30$ for ESO269-66. The RGB is thus partly broadened by photometric errors in the observed CMDs. However, the broadening does not come entirely from the errors, and thus is also associated with the physical properties of the galaxies. In particular, the color spread across the RGB could either be due to age or metallicity. There is a well known degeneracy in this evolutionary stage, such that stars that are younger and more metal--rich may be found in the same RGB region as stars that are older and more metal--poor. Ideally, for a ``simple'' stellar population a spread in age would produce a narrower RGB at the same metallicity than would a spread in metallicity at a constant age (see, for example, \citealt{vandenberg06}). Owing to the relatively small number of luminous AGB stars above the TRGB, we may assume that the majority of the RGB stars belong to old populations ($10$ Gyr or older). Because of this, and also because early--type dwarfs in the Local Group all display metallicity spreads, we assume that the color spread across the RGB is predominantly caused by a metallicity range within these galaxies. Our goal is then to derive photometric metallicity distribution functions from the CMDs.

We want to stress that a spread in metallicity would imply the presence of an age spread as well (of the order of $\sim1-2$ Gyr), such that the first generation of stars born in the galaxy would have time to evolve and pollute the surrounding interstellar medium, making it possible for the next star formation episodes to produce more metal--rich stars. This spread at such old ages is however not resolvable from the upper part of the RGB alone, and from our data we can only try to put some constraints on the age range of the intermediate--age populations from the number and brightness of the luminous AGB stars. The unresolvable spread in age for old populations could have the effect of inflating the derived metallicity spreads by $10-20\%$. On the other side, the possible presence of intermediate--age stars would also affect the metallicity spreads in a way that depends on each individual galaxy's characteristics, and it deserves more careful attention. Both these aspects will be discussed in detail in the next Section.


\section{Metallicity distribution functions} \label{mdf_sec}

\begin{figure*}
 \centering
  \includegraphics[width=13cm]{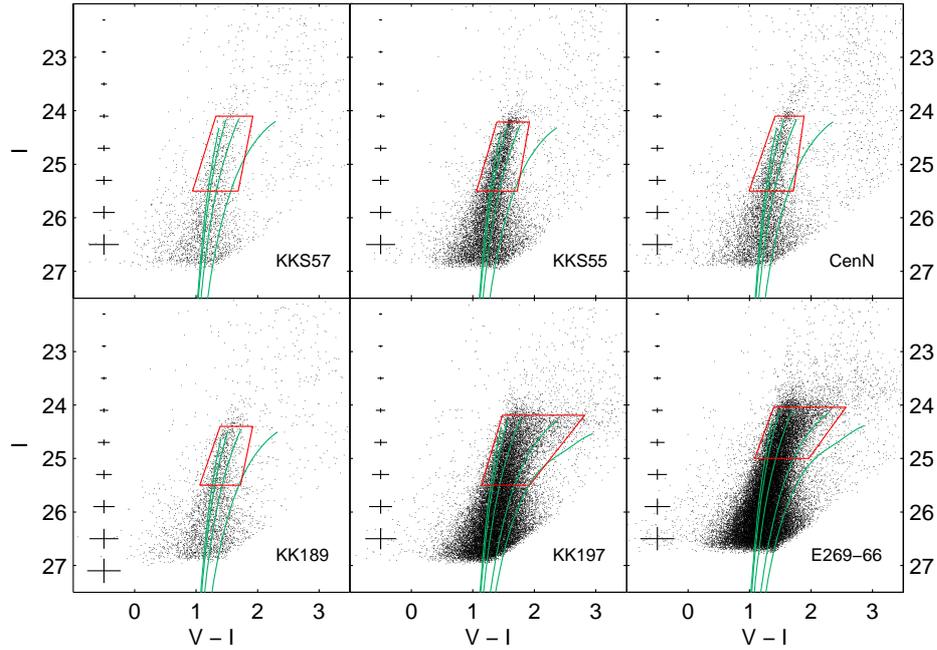}
 \caption{\footnotesize{Color magnitude diagrams of the six early--type dwarf galaxies studied in this paper, ordered by absolute $V$ magnitude of the galaxies, as in Fig. \ref{cmds}. Overlaid are Dartmouth stellar isochrones (green solid lines) with a fixed age of 10 Gyr, shifted to the distance of the galaxies and reddened according to the values listed in Table \ref{infogen}. Their metallicities have values of [Fe/H] $=-2.5$, $-1.9$, $-1.3$ and $-0.7$ (proceeding from the blue to the red part of the CMD). For KK197 and ESO269-66, also the isochrone with [Fe/H] $=-0.4$ is shown, because of the broader RGB. Drawn in red is the selection box within which we interpolate metallicity values for the single RGB stars (see text for details).}}
 \label{cmds+iso}
\end{figure*}

\begin{figure*}
 \centering
  \includegraphics[width=14cm]{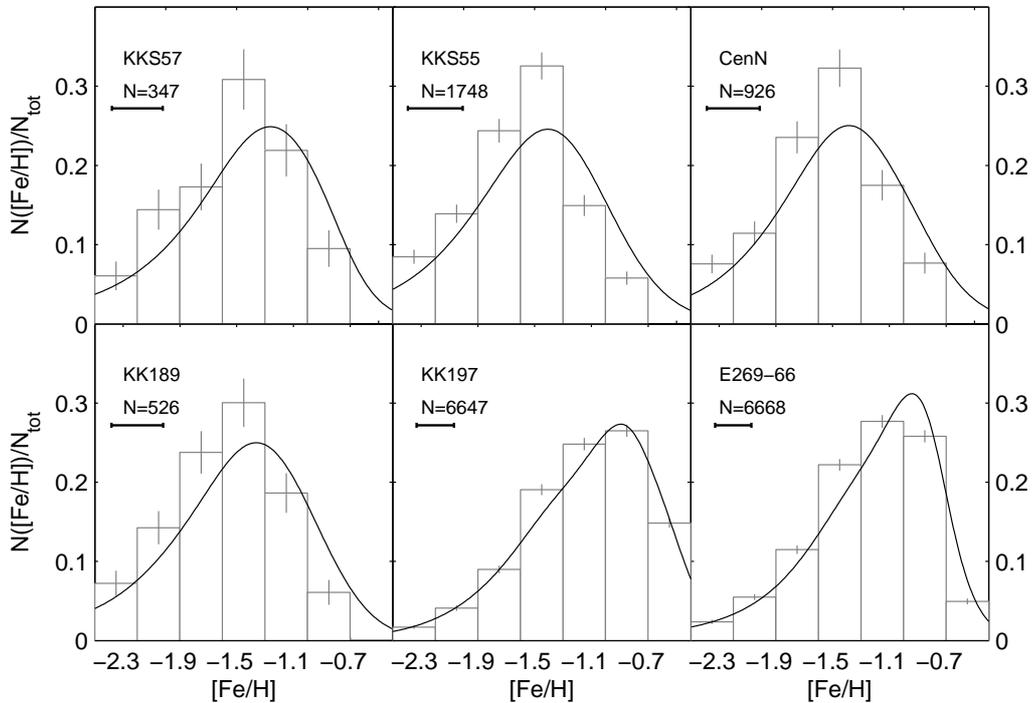}
 \caption{\footnotesize{Normalized metallicity distribution functions of the six dwarfs, derived via interpolation of isochrones at a fixed age (10 Gyr) and varying metallicity. Overlaid (black lines) are the MDFs convolved with the observational errors. Also plotted in the left upper corner is the median error on the individual values of [Fe/H], and the total number of stars considered (for details see Sect. \ref{mdf_sec}).}}
 \label{mdfs}
\end{figure*}

\begin{table*}
 \centering
 \caption{\footnotesize{Metallicity values derived for the studied sample of galaxies}} 
 \label{infores}
\begin{tabular}{lcccccccc}
\hline
\hline
Galaxy&$N_{RGB}$&$\%_{cont}$&$<$[Fe/H]$>_{med}$&$\sigma$[Fe/H]$,obs$&$\sigma$[Fe/H]$,int$\\
\hline
\object{KKs57}&$347$&$9$&$-1.45$&$0.46$&$0.28$\\
\object{KKs55}&$1748$&$2$&$-1.56$&$0.40$&$0.1$\\
\object{CenN}&$926$&$5$&$-1.49$&$0.40$&$0.15$\\
\object{KK189}&$526$&$5$&$-1.52$&$0.42$&$0.2$\\
\object{KK197}&$6647$&$1$&$-1.08$&$0.49$&$0.41$\\
\object{ESO269-66, KK190}&$6668$&$0.5$&$-1.21$&$0.42$&$0.33$\\
\hline
\end{tabular}
\end{table*}

At the distance of the Centaurus A group ($\sim3.8$ Mpc on average, \citealt{kara07}), there is no possibility to obtain individual stellar absorption line spectra for red giants using the current instrumentation within reasonable integration times (compare $I_{TRGB}$ in Table \ref{infogen}). Moreover, genuine early--type dwarf galaxies usually do not contain neutral gas reservoirs nor do they show ongoing star formation. Hence, we cannot directly measure their present-day metallicity from spectroscopy of HII regions around massive young stars. Finally, a reconstruction of the star formation history from CMDs would be almost impossible, given the small amount of information that comes from the RGBs only.

The only tool we have to constrain metallicity in this type of galaxies and at these distances, is thus photometry combined with isochrones. Once we have a set of isochrones from an evolutionary model, we can overlay them on the observed CMDs, as shown in Fig. \ref{cmds+iso}. In the case of a single stellar population with a narrow RGB, one would try to find the one best--fitting isochrone. However, early--type dwarf galaxies, at least in the Local Group, are known to have large spreads in their metallicities (e.g., \citealt{grebel97}; \citealt{mateo98}). As our RGBs look quite broad and there are no young population features in the CMDs (i.e., stars younger than $\sim1$ Gyr, as argued in the previous Sect.), we make the simplified assumption of a single, fixed old age and then let the metallicity of the isochrones vary to cover the whole RGB color range. In this way we are able to derive the metallicity of each star on the RGB via interpolation among the isochrones. This method is widely used for studies of predominantely old populations, for which spectroscopy is not available \citep[e.g., ][]{durrell01, sara02, rejkuba05, harris07, richardson09}, to derive photometric metallicity distribution functions.

In this method, there are some weaknesses that have to be taken into account. Firstly, we will have a small ($\sim22\%$, see e.g. \citealt{durrell01}) contamination from old ($\gtrsim10$ Gyr), low-luminosity AGB stars which overlap with our RGB. With the available photometry it is impossible to distinguish them from RGB stars, and our main conclusions are not affected by this inevitable contamination since it would introduce a systematic bias towards slightly lower metallicities within the investigated metallicity range. That is, we would have an approximately equally overestimated numer of stars for each metallicity bin. Secondly, we observe the presence of luminous AGB stars above the TRGB, which resemble an intermediate--age, metal--poor population (going up almost straight to higher luminosities), meaning that in the most metal--poor bins of our resulting metallicity distribution functions there will be a non-negligible contribution from these stars. This effect has to be quantified more accurately for each galaxy, and will be considered in the next Section.

In Fig. \ref{cmds+iso} we overplot on the observed CMDs the boxes (in red) used to select the putative stars for which we derive metallicities. Their width in color is chosen by constructing Hess diagrams for the galaxies, so as to approximately contain stars that lie within $\sim\pm3\sigma$ from the mean locus of the RGB. Their vertical size is such that the stars in the selection boxes have $1 \sigma$ photometric errors of $\leq0.1$ mag in magnitude and $\leq0.15$ mag in color, and go up to the TRGB (the apparent magnitude of the latter has been adopted from \citealt{kara07}). In the selected region, the $I$-band completeness is above the $\sim70\%$ level and the theoretical isochrones are more widely separated from each other than at lower luminosities, providing a better resolution in metallicity.

In Fig. \ref{cmds+iso} we further show the adopted stellar isochrones, taken from the Dartmouth evolutionary models \citep{dotter08}. We chose not to use $\alpha$-enhanced tracks, because nothing is known about the level of the $\alpha$-enhancements in our target galaxies. The effects of this arbitrary choice will be discussed in the next Section. The isochrones are shifted to the distance of each observed galaxy and reddened by the respective foreground reddening value. Distances and reddening values are taken from \citet{kara07} and from the Schlegel extinction maps \citep{schlegel98}, respectively. The Dartmouth set of evolutionary models is able to reproduce particularly well the populations of old and intermediate--age clusters, while other models generally fail to simultaneously reproduce all features of the CMD for the correct, spectroscopically measured metallicity (e.g., \citealt{glatt08a}; \citealt{glatt08b}; \citealt{sara09}). We choose a fixed age of 10 Gyr, and metallicities ranging from [Fe/H] $=-2.5$ to $-0.3$ in solar units. The implications of this simplistic, single age assumption are discussed in detail in the next Section. The isochrone grid is finely spaced ($0.2$ dex steps), in order to get good interpolation values for each star. We interpolate linearly among the isochrones. We also only retain stars that fall within the range of our isochrone grid, while rejecting those with extrapolated values (meaning values blueward of the most metal--poor isochrone and redder than the most metal--rich one), to avoid artificially metal--poor or --rich extremes in the results. 

After deriving the metallicity values for each RGB star through interpolation, we show in Fig. \ref{mdfs} the metallicity distribution functions (MDFs) for our galaxies (each normalized to the total number of considered stars). The galaxies are again ordered by increasing absolute $V$ magnitude. We choose not to correct the MDFs for completeness, since the latter changes very little within the magnitude range of the selection box. The errors in [Fe/H] are derived as follows: we perform 1000 Monte Carlo realizations of the interpolation process, varying the position of the stars on the CMD within their respective photometric errors (assuming Gaussian distributions), thus accounting for random errors. The median $1\sigma$ measurement errors of the individual values of [Fe/H] are plotted in the left corner of each subpanel. They range from $\sim0.26$ to $\sim0.39$ dex. The vertical (random) counting errors of each histogram bin are also taken from the Monte Carlo realizations. Finally, in the figure we also draw the MDFs convolved with observational random errors (black lines). The effects of the systematic errors (on distance and reddening of the galaxy, taken from \citealt{kara07}) on the MDFs will be discussed in the next Section.

All six galaxies contain populations that are on average metal--poor, with median values ranging from $<$[Fe/H]$>_{med}=-1.56$ to $-1.08$ (the median value is more meaningful since the MDFs are not symmetric). There is no galaxy that appears to contain stars with a metallicity higher than [Fe/H] $=-0.4$. This result is influenced by our choice of the selection box, but for clarity in Fig. \ref{cmds} we overplot (for the galaxies KK197 and ESO269-66) also an isochrone with [Fe/H] $=-0.4$: indeed, it can be seen how few stars lie redwards of this isochrone, so we conclude that our cutoff is reasonable. On the other side, the few stars that lie bluewards of the bluest isochrone (see Fig. \ref{cmds+iso}) have probably been scattered there by photometric errors. Some of them may also be old AGB stars and some of them may be genuinely metal--poor RGB stars. We have no way of identifying the nature of such stars in our photometry. Hence, as stated earlier, we simply avoid extrapolating toward putative metallicities lower than covered by our isochrones.

The spreads in metallicity are quite large. The nominal ranges of the metallicity covered in each galaxy are $\Delta$[Fe/H]$,obs\sim2$ dex, and reach typically from [Fe/H] $=-2.5$ to values as high as [Fe/H] $=-0.7$ to $-0.4$. In the classical dwarf spheroidals in the Local Group the full spectroscopically measured RGB metallicity range typically well exceeds 1 dex (e.g., \citealt{grebel03}), so this large range is in good agreement with what one might expect. When calculating a formal Gaussian dispersion the metallicity spreads derived here are of the order of $\sigma$[Fe/H]$,obs\sim0.4$--$0.5$ dex. This result requires some further attention. That is, we have to take into account the median error on the individual metallicity estimates, which comes from both photometric errors and from the close spacing of the isochrones on the CMDs. This simply means that the true metallicity dispersion is in fact narrower than derived. For a better estimate of the global intrinsic spread of the galaxy, $\sigma$[Fe/H]$,int$, we thus subtract in quadrature the median metallicity error from the observed dispersion. The final values range from $\sigma$[Fe/H]$,int\sim0.10$ dex to $\sim0.41$ dex. This implies that the observed broadening of the RGBs is not only an effect of photometric errors.

In Table \ref{infores} we summarize the results as follows: column (1): name of the galaxy, (2): number of RGB stars for which we derive individual metallicity values, (3): percentage of Galactic foreground contaminants with respect to the number of stars in the selection box (see next Sect.), (4): median [Fe/H] value computed from the MDF, (5): observed metallicity dispersion of the MDF (from Gaussian fit), and (6): intrinsic metallicity dispersion, after subtraction of the median measurement error.

The six histograms shown in Fig. \ref{mdfs} show a slow decline in the metal--poor direction, and a steeper slope on the metal--rich side. The results found here are similar to what has been observed in dwarf galaxies of the Local Group, both in terms of metallicity ranges and of MDF shape \citep[e.g.,][]{shetrone01, sara02, tolstoy04, battaglia06, helmi06, koch06, bosler07, koch07a, koch07b, gullieu09}. However, since in our work we are only able to present results based on limited assumptions because of the age--metallicity degeneracy, we do not compare the shapes of our MDFs to theoretical models or to Local Group members in more detail. We just note that the overall shapes of the MDFs resemble the MDFs of Galactic dwarf spheroidals, which suggests that similar evolutionary processes may have governed their star formation and enrichment histories.


\section{Possible sources of error} \label{serrors}

The results presented in Tab. \ref{infores} will unavoidably be affected by the assumptions we made in the first place, i.e. a single old age, a negligible contribution to the MDFs by intermediate--age stars, and a scaled solar value for the $\alpha$-element abundances. We discuss now more in detail the effects that these assumptions may have on the derived results.

\begin{figure}
 \centering
  {\includegraphics[width=7.cm]{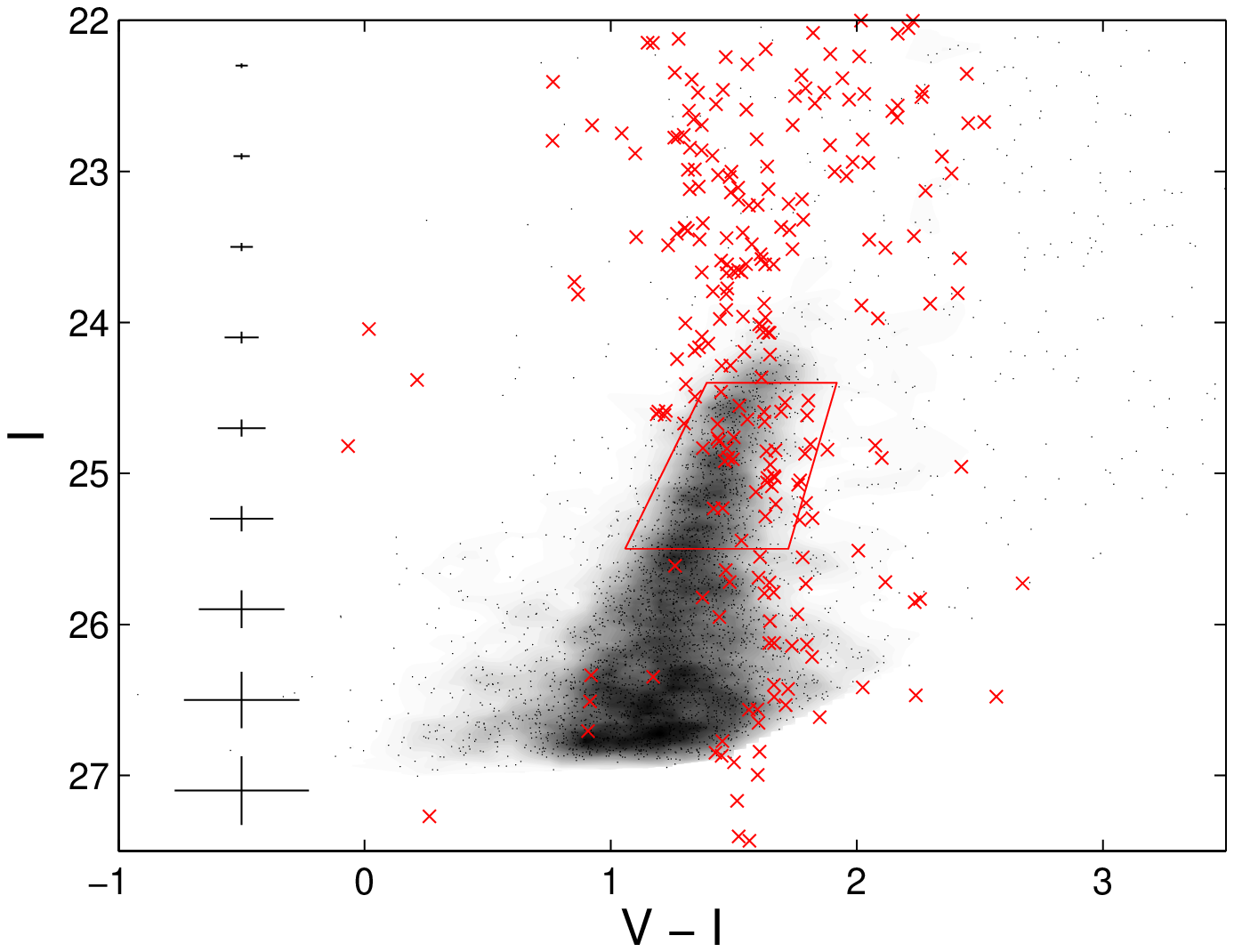}
  \includegraphics[width=7.cm]{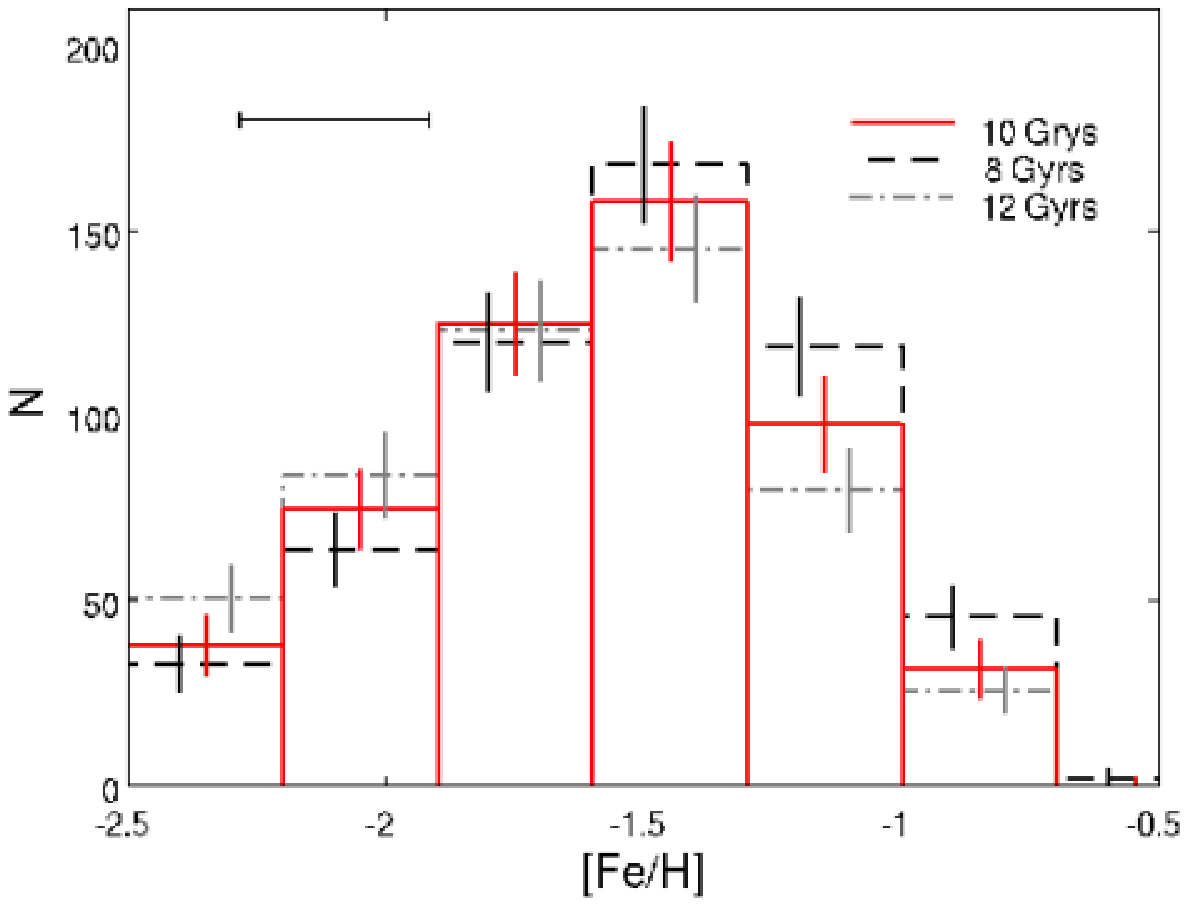}
  \includegraphics[width=7.cm]{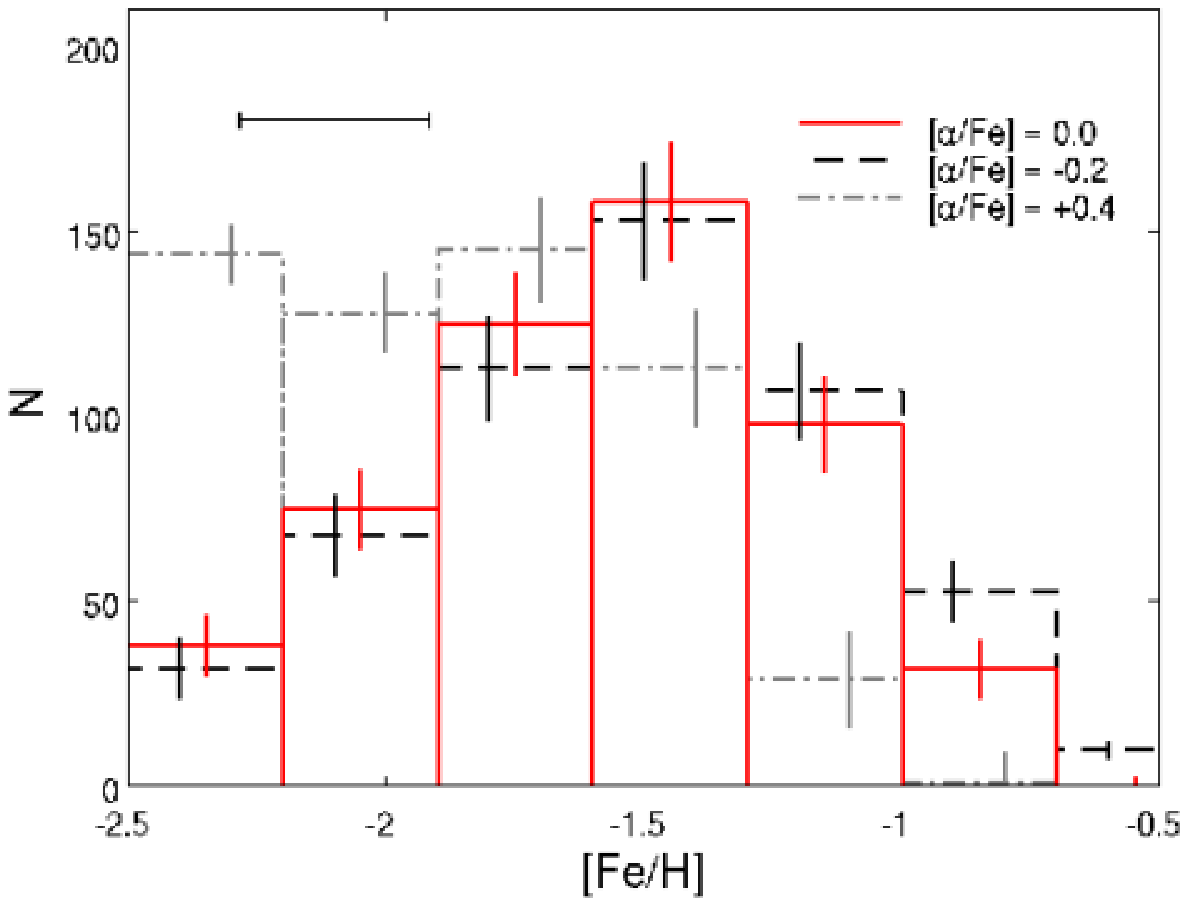}}
 \caption{\footnotesize{\emph{Upper panel.} Color magnitude diagram of the dwarf spheroidal galaxy KK189, overlaid by a Hess density diagram of its stars. Shown also is the expected location of Galactic foreground stars (red crosses), estimated from the TRILEGAL models, and the selection box within which we derived metallicities. \emph{Central panel.} Metallicity distribution function of KK189, derived via interpolation of isochrones at a fixed age and varying metallicity. The results are shown for 3 different ages (8, 10 and 12 Gry), and the errorbars are computed considering photometric errors. Also plotted in the left corner is the median individual metallicity error. \emph{Lower panel.} As for the central panel, but keeping the age fixed at 10 Gyr and varying the $\alpha$-enhancement abundance ([$\alpha$/Fe]$=-0.2,+0.0$ and $+0.4$).}}
 \label{errors}
\end{figure}

\subsection{Foreground contamination}

For the HST/ACS data used here, we do not expect a high number of unresolved background galaxies in such a small field of view. In fact, they will be rejected by the DOLPHOT quality cuts, leaving a contamination of the CMDs of less than $3\%$ \citep{dalcanton09}.

Regarding Galactic foreground contamination, we obtain estimates using the TRILEGAL models \citep{girardi05}. In the direction of the Centaurus A group, and within the entire ACS field of view, there are on average $\sim200$ foreground stars expected. Additionally, the TRILEGAL models provide the photometry of the simulated stars, so that we are able to check which regions of the CMD would be most affected by foreground stars. In particular, we compute the percentage of contaminant stars found in the RGB selection boxes for each galaxy, after taking into account completeness effects (not all of the foreground stars would in fact be detected by the instrument), and we report the values in Table \ref{infores}. They range from $<1\%$ to $\sim9\%$, depending on the total number of stars in each galaxy.

In the upper panel of Fig. \ref{errors} we take the galaxy KK189 as an example: we plot a Hess density diagram of its stars on top of the CMD to illustrate more clearly its main features, and overplot the expected location of Galactic foreground stars (red crosses) derived from the TRILEGAL models. The predicted contaminants are few and are relatively uniformely distributed within the selection box, so we decided not to statistically subtract them from our RGB sample. Given the errorbars on the MDFs, the small number of stars that would be subtracted in such a procedure would indeed not change significantly our final results. With respect to the density maps presented in Sect. \ref{ssd_sec}, the lowest contour level is always well above the value inferred for foreground contamination.

It could be argued that the Galactic models are not always perfectly able to reproduce the observed contamination. In particular, they underestimate the amount of objects with colors $V-I>2$, partly because the contribution from thick disk and halo stars is uncertain at these colors and partly because there could be unresolved galaxies among them. However, the predicted star counts will differ in different color bins as a function of the slope of the adopted Galactic IMF. Thus, the underestimation at red colors does not necessarily mean that the models are uncorrectly reproducing the contamination at the position of our RGBs.

From Fig. \ref{errors} we also see that the contamination is higher in percentage for the upper part of the CMD above the TRGB, since the number of luminous AGB stars is very low. It is difficult from the optical observations alone to thus clearly distinguish between real AGB stars and foreground contaminants. This point will be considered again below, where we investigate more in detail the contribution from intermediate--age objects.

We also investigate the possible foreground contamination by projected halo stars of the dominant galaxy Centaurus A. This contamination may happen for the dwarfs that are found at very close projected distances from the giant elliptical, as in the case of KKS55 and KK197 ($\sim48$ and $\sim58$ kpc, respectively). The RGB of Centaurus A has been investigated in detail in a series of papers \citep{harris99, harris00, harris02, rejkuba05} in order to derive the MDFs for fields at different galactocentric distances (8, 21, 31 and 38 kpc). If we assume that the halo of the giant elliptical extends further, out to the projected distances at which KKS55 and KK197 are found, and that the MDF shape of Centaurus A will not be much different from the more internal ones at these distances, we may then try to look for a possible contamination in the CMDs of our dwarfs. The photometric data collected for Centaurus A show a range in magnitude and color that is very similar to those of the dwarfs in our study. Thus, if the contamination from the Centaurus A halo stars is relevant, we should tentatively observe the RGB features of the giant elliptical even in our CMDs, i.e. we should see a very broad RGB with an extended metal--rich population (the peak metallicity for the outermost field of the galaxy is of [Fe/H]$\sim-0.6$). This is not found for the closest dwarf in projection, KKS55, which displays a quite narrow RGB, and so we do not expect it to be true either for KK197. For the sake of completeness, we check the radial distribution of the most metal--rich stars found in the CMD of KK197, and found them to be centrally concentrated, thus excluding a possible, diffuse contamination from the Centaurus A field.

\subsection{Age assumption}

In our metallicity derivation, we may be neglecting the presence of stellar populations younger than the adopted age of 10 Gyr. In the CMDs of our targets (Fig. \ref{cmds}) some luminous AGB stars are indeed visible above the TRGB. The RGB could be contaminated by the stars coming from these intermediate--age populations that are on the RGB or ascending to the AGB phase, after having burned off the He in their cores. This contamination will be mostly in the metal--poor part of the derived MDFs, as relatively metal--poor, intermediate--age stars in their RGB/AGB stage would overlap the old, most metal--poor isochrones (age--metallicity degeneracy).

We give a rough estimate of the number of luminous AGB stars by considering Padova stellar evolutionary models \citep{marigo08}, since the Dartmouth isochrones do not model stages later than the RGB phase. We select stars in a box that extends (in magnitude) from $\sim0.1$ mag above the TRGB (this is greater than the photometric error in magnitude at these luminosities, so we make sure we are not picking RGB stars upscattered by photometric errors) up to $\sim1.1$ mag above the TRGB. The range in color goes from the bluest edge of the RGB to $V-I\sim3$. This selection criterion should still retain luminous AGB stars with ages from 1 to 10 Gyr, and with metallicities that are around the median metallicities derived for our galaxies (we assume that the intermediate--age populations will be on average slightly more enriched than the oldest and most metal-poor RGB stars).

We first compute the bolometric magnitudes for the candidate AGB stars with the empirical correction formula by \cite{dacosta90} (for $I$-band magnitude). We then construct the bolometric luminosity function for the AGB stars, subtracting the number of predicted foreground stars for each magnitude bin. Since the star counts from the TRILEGAL models could be uncertain, we also try a second method, subtracting the number of stars that are found in a box with the same size of the AGB selection box, but just above it on the CMD. The two methods give results that are almost identical.

Having done this, we can now use the empirical relation derived in \cite{rejkuba06} (their Fig. 19), which connects the tip luminosity of the AGB and their age (for ages $\geq 1$ Gyr). This method gives us a rough quantitative idea about the fraction of an intermediate--age population. Finally, we use an approximation for the formula of the fuel consumption theorem (\citealt{armand93}, originally from \citealt{renzini86}) to compute the expected number of luminous AGB stars per magnitude, given their fraction relative to the entire galaxy population. We then compare the results from our AGB counting to the predictions from this theorem, varying the fraction of the intermediate--age contaminants until the two values are similar.

We now obtain for each galaxy an estimate of the presence of stars with ages in the range $\sim4-8$ Gyr. Younger stars are probably not present as seen from our data, if we assume our foreground subtraction is correct. The results are as following: for KK189, KKS57 and KK197 the estimated fraction of the intermediate--age population relative to the entire population is of $\sim10\%$; ESO269-66 has a probable fraction of $\sim15\%$, while for KKS55 and CenN this number grows up to $\sim20\%$. All of the stated fractions could possibly be lower limits to the true values, since the luminous AGB phase is short and thus this stage could be poorly populated in the observed CMDs. These numbers are lower than what is observed in those few Local Group early--type dwarf galaxies with pronounced intermediate--age star formation (e.g., Leo I, Leo II, Fornax, Carina, NGC147 and NGC185, with fractions up to $\sim50\%$), but in line with previous studies of other early--type dwarfs in the Centaurus A group \citep{rejkuba06}.

At this point, in our MDFs we simply subtract the fraction of intermediate--age stars from the metallicity bins that are more metal--poor than the median metallicity. If we then recompute median metallicities and spreads for each galaxy, we find that: for KK189, KKS57 and KK197, the median metallicities values would not change significantly since the amount of intermediate--age stars is small, and the spreads would be slightly narrower ($\sim5-10\%$); for ESO269-66 the new median metallicity is higher by $\sim5\%$ and the spread is smaller by $\sim10\%$; finally, KKS55 and CenN would be only slightly more metal--rich ($\sim2-3\%$), but the Gaussian metallicity spreads would practically go to zero.

However, we want to underline that the best way to look more in detail at luminous intermediate--age stars is rather to combine optical with near-infrared filters, because they are more luminous in the infrared and more easily separated from foreground contamination in the latter ones (e.g., \citealt{rejkuba06}; \citealt{boyer09}). We will consider ground based (Very Large Telescope) images in $J$- and $K$-band in a companion paper (Crnojevi\'{c} et al. 2010, in prep.), in order to get better constraints on the epoch and strength of the more recent star formation episodes for our target galaxies.

The assumption of one single age for the galaxies' populations is also a strong simplification. In order to account for the spread in metallicity, extended star formation histories are needed. For early--type dwarf galaxies in the Local Group, complex star formation histories have been derived, and it was shown that no two dwarfs are alike, not even within the same morphological subtype \citep{grebel97}. Although there are examples of Local Group dwarf spheroidals with predominantely old and metal--poor populations (e.g., Ursa Minor, Draco), the case of a single, ancient starburst is quite unlikely for most of them, on the contrary large metallicity spreads have been detected, and the early star formation appears to have been low and continuous (e.g., \citealt{ikuta02}; \citealt{grebel03}; \citealt{koch06}; \citealt{coleman08}; \citealt{koleva09}). Different models arrive at different durations to explain the observed metallicity spread (e.g., \citealt{ikuta02, lanfranchi04, marcolini08}). Assuming that the spread in metallicity is predominantly due to a spread in age, we repeated the metallicity interpolation process choosing the three most metal--poor isochrones to have an age of 12 Gyr, and the most metal--rich ones an age of 8 Gyr. The result of this simple test is that, as one would expect, the peak of the MDF is less pronounced, and the intrinsic dispersion is slightly larger ($\sim15\%$ in dex). However, since the observations are not deep enough to permit us to resolve the age--metallicity degeneracy from photometry of main sequence turnoff stars, a single age is the simplest assumption we can make without going into pure speculation. Moreover, as mentioned earlier, metallicity is the main contributor to the shape and width of the RGB; age has much less of an effect.

Having clarified this, our choice to set the isochrone age to 10 Gyr is arbitrary, but as we can see from the central panel of Fig. \ref{errors} we may make such an assumption. We plot three different MDFs for KK189: one is derived from our chosen age, the other two are derived from ages slightly lower ($8$ Gyr) and slightly higher ($12$ Gyr) than the chosen one. It is clear that we do not introduce a significant bias when choosing one isochrone age instead of a slightly older or slightly younger one. Only in the case of an assumed younger age, there will be a higher number of metal--rich stars, but this will not significantly change the median metallicity value (the amount of change is $\sim5\%$ in dex), nor the shape of the distribution. The reason for this relatively small dependence on age is that a decrease or an increase in age by a few Gyr has very little effect on the isochrones for these high ages. Hence a change in metallicity is much more noticeable, as it will have a larger effect on the isochrones.

\subsection{$\alpha$-enhancement}

For our adopted set of isochrones, we chose not to use $\alpha$-enhanced tracks, because nothing is known about the level of the $\alpha$-enhancements in our target galaxies. In the Local Group, the dwarf spheroidals present a broad range of [$\alpha$/Fe] ratios at the metallicities of our galaxy sample \citep[e.g., ][]{shetrone01, shetrone03, sadakane04, geisler07, koch08}. In particular they can vary from sub- ([$\alpha$/Fe]$\sim-0.2$) to super-solar ([$\alpha$/Fe]$\sim+0.4$) values within the same dwarf, and the observed [$\alpha$/Fe] ratios show a correlation with the [Fe/H] ratios, such that the $\alpha$-element abundance tend to decrease for increasing [Fe/H] (e.g., \citealt{tolstoy09, koch09}).

We try again the simple test described above for a choice of different ages, and let the $\alpha$-element abundances vary for our isochrones. With a fixed age of 10 Gyrs, we re-compute the metallicities for isochrones with [$\alpha$/Fe]$=-0.2,+0.2$ and $+0.4$. The resulting MDFs are shown in the lower panel of Fig. \ref{errors} for [$\alpha$/Fe]$=-0.2,+0.0$ and $+0.4$. Changing the $\alpha$-element abundances by $\pm0.2$ from the solar values gives a similar result as a change in age, such that lower (higher) $\alpha$-element abundances would mimic a younger (older) set of isochrones, and the median metallicities would be accordingly higher (lower) by less than $\sim5\%$. When choosing [$\alpha$/Fe]$=+0.4$, the median metallicity is lower by $\sim30\%$, and the metallicity spread is almost doubled. However, this is a rather extreme assumption, since the most probable case is the one where we have a spread of $\alpha$-enhancement values. This could again be included in our computation by choosing the most metal--poor and oldest isochrones to have sub-solar $\alpha$-enhancement, and the most metal--rich and youngest to have super-solar $\alpha$-enhancement. Our simple test is primarly intended to explore the parameter range. It demonstrates that the resulting median metallicities would be slightly lower ($\sim15\%$ in dex), while the metallicity spread would be much larger (3-4 times) when a range of [$\alpha$/Fe] values is considered.

\subsection{Systematic errors}

In our analysis, we adopt the values for distance and foreground reddening from \citet{kara07} and from the Schlegel extinction maps \citep{schlegel98}, respectively. These are affected by uncertainties of the order of $\sim7\%$ and $\sim10\%$ of the values listed in Tab. \ref{infogen}. If we let one of these two parameters vary within its errorbars, the isochrones overplotted on the observed CMDs will be all shifted by the same amount.

We first test how the derived median metallicities and metallicities spreads change if we change the adopted distance value within its errorbars. We find that for distances higher (lower) than the adopted one, the median metallicities are lower (higher) by $\sim10\%$ (meaning $\sim0.10-0.15$ dex), while the spreads are also higher (lower) by $\sim20\%$. Similarly, for higher (lower) foreground reddening values the resulting median metallicities will be higher (lower) by $\sim3\%$ and the spreads lower (higher) by $\sim10\%$.

\subsection{Stellar evolutionary models}

As a test, we also repeated our entire analysis using Padova isochrones \citep{marigo08}, with an identical age and metallicity grid. The results are overall quite similar, from the shape of the MDFs to the presence of metallicity gradients and distinct stellar populations (see next Sect.), but the average metallicities found for each galaxy are systematically higher by $\sim0.3$ dex on average. However, a detailed comparison between different theoretical models goes beyond the goals of our work (for some examples, see \citealt{gallart05}; \citealt{glatt08a}; \citealt{glatt08b}; \citealt{goud09}; \citealt{sara09}).


\section{Metallicity and population gradients} \label{ssd_sec}

\begin{figure*}
 \centering
  \includegraphics[width=14cm]{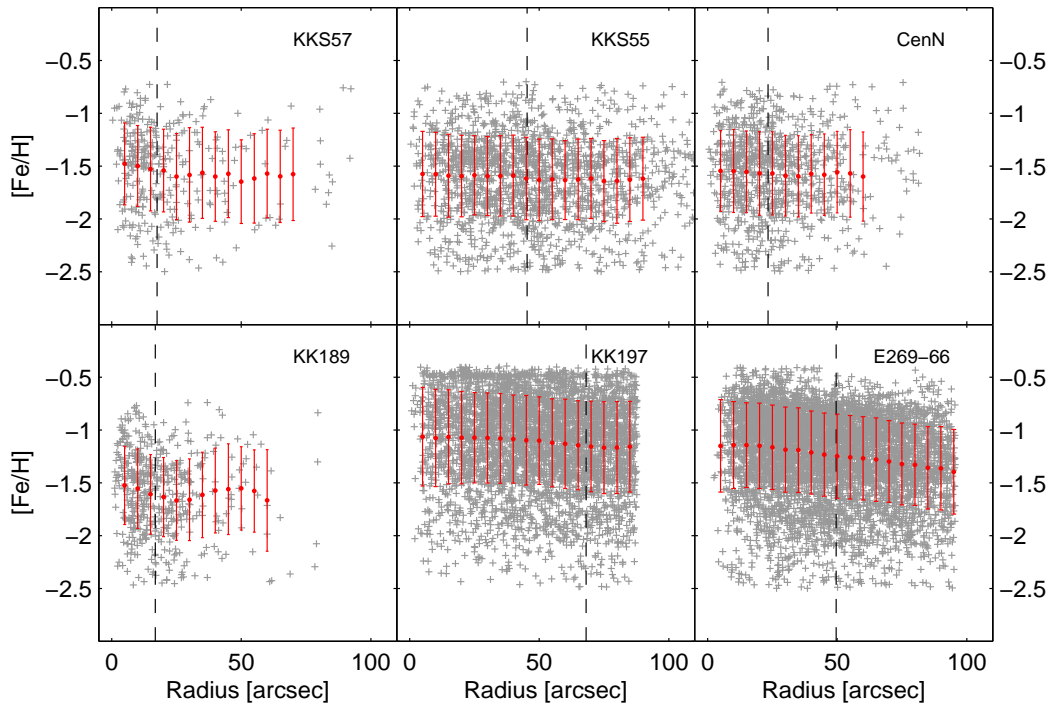}
 \caption{\footnotesize{Metallicities of individual RGB stars as a function of elliptical radius for the six galaxies. The projected half--light radius $r_{h}$ is plotted as a dotted line in each subpanel. A running mean (red points) is also drawn for each galaxy.}}
 \label{mgrad}
\end{figure*}

\begin{figure*}
 \centering
  \includegraphics[width=14cm]{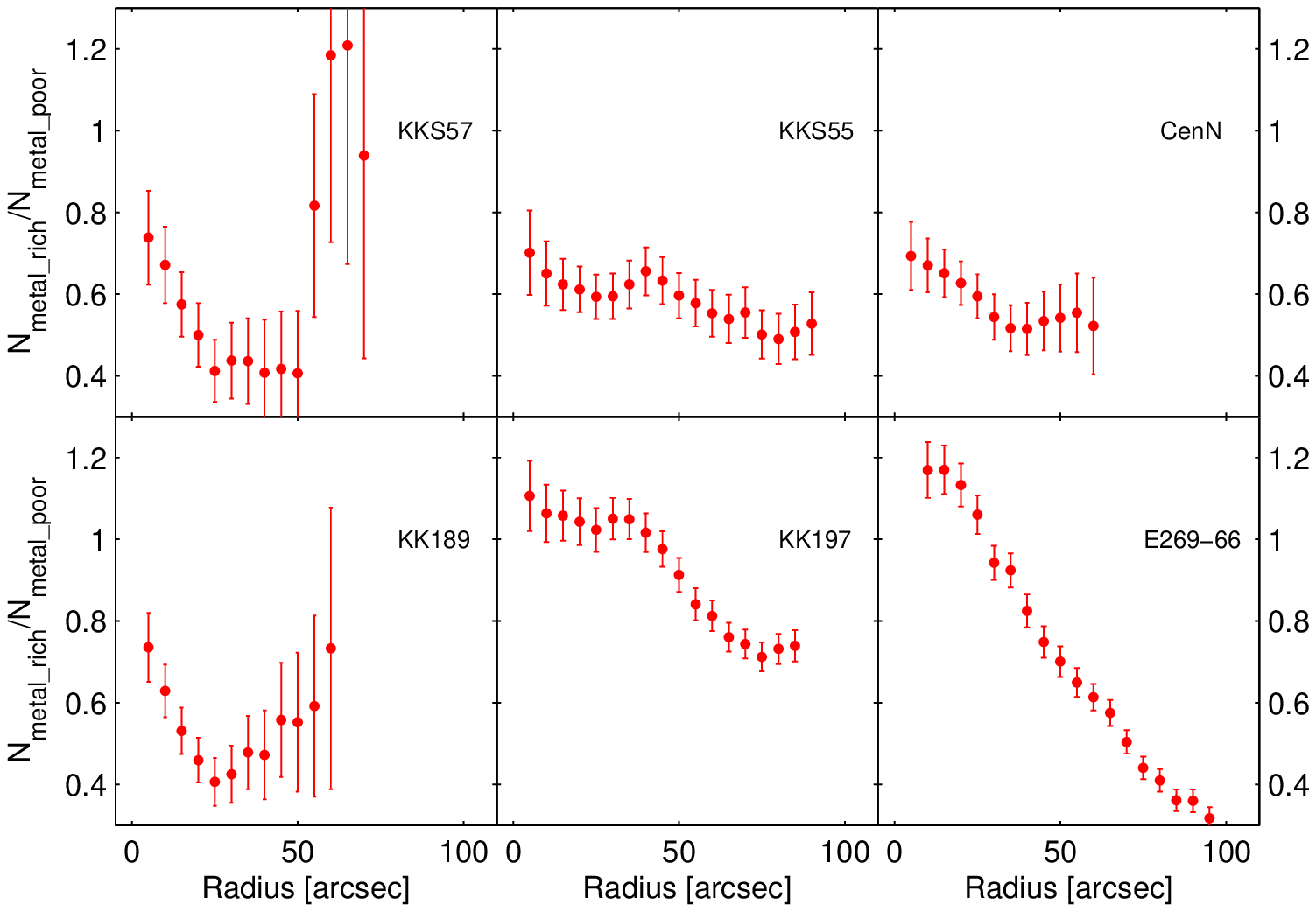}
 \caption{\footnotesize{Ratio of metal--rich to metal--poor stars as a function of elliptical radius. These stars have metallicities values $>(<$[Fe/H]$>_{med}+0.2)$ and $<(<$[Fe/H]$>_{med}-0.2)$, respectively.}}
 \label{mrtomp}
\end{figure*}

We now look for metallicity gradients within the galaxies, and for projected spatial variations in their stellar populations.

\subsection{Metallicity gradients?}

To take into account the fact that the galaxies do not have a perfectly circular shape, we consider elliptical radii from now on. The determination of the ellipticity for each galaxy follows the method of \citet{mclaug94}.  In Fig. \ref{mgrad} we show the individual metallicity values for the RGB stars as a function of elliptical radius. We plot the data until the last ellipse that is entirely contained within the ACS field of view, because data beyond that point are incomplete in metallicity. That is, we could correct the number of stars to account for the fact that we do not observe the entire area of the ellipse, but we could not have any metallicity information about the stars that lie outside the instrumental field of view. For the galaxies KKs55, KK197 and ESO269-66, the instrumental field of view does not cover their whole extent. We also overplot the projected half--light radius for each galaxy, derived as described below.

We compute projected, completeness corrected surface density profiles from our RGB stars, and fit them with 3-parameter King models \citep{king62}. However, these models yield only poor results when we attempt to simultaneously fit both the very central parts and the outskirts of the dwarf galaxies. We thus also derive $I$-band surface brightness profiles from star counts, and fit them with Sersic profiles \citep{sersic68}. This is a generalized exponential function of the form $I(r)\sim I(r_h)\exp[-(r/r_h)^{1/n}]$, where the surface brightness is expressed in terms of intensity. It should be noted that for this type of parametrization some authors use the index $n$ instead of ${1/n}$. The free parameters for the Sersic profile are the effective (half--light) radius $r_{h}$, a shape parameter $n$ that characterises the curvature of the profile, and the surface brightness at the radius $r_{h}$. The values derived here are reported in Table \ref{sersic}. We cannot make direct comparisons with the values previously published for our sample of galaxies because such studies consider surface brightness profiles from different bands (\citealt{jerjen00b}), or perform the fit with a simple exponential profile (\citealt{sharina08}). The results derived in this work are nevertheless consistent with the ones found in the aforementioned studies. For the sake of clarity, we have to underline that we construct our surface density profiles as a function of elliptical radius. However, other authors often define a ``reduced radius'' ($r_{red}=\sqrt {ab}$, with $a$ and $b$ as the semi-major and semi-minor axis, respectively) to account for the ellipticity of the galaxy. In the case of a small ellipticity, the two radii are consistent with each other, but in the case of a high ellipticity they can differ substantially. In our sample, the galaxy KK197 is very elongated so its ``reduced'' half--light radius would result in a value smaller than the one computed from our profiles. For completeness, we apply the correction for ellipticity to our derived half--light radii and report them as well in Table \ref{sersic} ($r_{h,red}$). We will use this information later, when looking for different stellar populations within our galaxies.

\begin{table}
 \centering
 \caption{\footnotesize{Parameters from Sersic profiles}} 
 \label{sersic}
\begin{tabular}{lccc}
\hline
\hline
Galaxy&$r_{h}^{\mathrm{a}}$&$r_{h,red}^{\mathrm{b}}$&$n$\\
\hline
\object{KKs57}&$17.5\pm0.8$&$14.2\pm0.6$&$0.97\pm0.14$\\
\object{KKs55}&$45.3\pm0.6$&$43.9\pm0.6$&$0.79\pm0.04$\\
\object{CenN}&$23.3\pm0.6$&$22.5\pm0.5$&$0.73\pm0.05$\\
\object{KK189}&$16.8\pm0.5$&$14.9\pm0.4$&$0.88\pm0.06$\\
\object{KK197}&$68.1\pm1.5$&$47.2\pm1.0$&$0.60\pm0.05$\\
\object{ESO269-66, KK190}&$49.6\pm0.6$&$43.6\pm0.5$&$0.76\pm0.04$\\
\hline
\end{tabular}
\begin{list}{}{}
\item[$^{\mathrm{a}}$] computed from surface brightness as a function of elliptical radius.
\item[$^{\mathrm{b}}$] computed from surface brightness as a function of ``reduced'' radius $r_{red}=\sqrt {ab}$ (see text for details).
\end{list}
\end{table}

Coming back to Fig. \ref{mgrad}, it is not easy to see a possible trend in such plots, thus we also compute and plot a running mean for each galaxy (in steps of $\sim5$ arcsec). Simple linear fits reveal flat or weak overall gradients. For KKS55, KK197 and ESO269-66, the results are $\sim-0.043$, $-0.09$ and $-0.17$ dex per arcmin respectively (or $\sim-0.036$, $-0.075$ and $-0.15$ dex per kpc). For the three remaining galaxies, which are less extended than the ones above, a weak gradient is observed in the central regions, while in the outer parts it tends to flatten (and the statistic tends to worsen because of the small number of stars in the outskirts). We find that: for KKS57, within the inner $\sim40$ arcsec, we have a gradient of $\sim-0.18$ dex per arcmin ($\sim-0.15$ dex per kpc); for CenN, the value is of $\sim-0.078$ dex per arcmin ($\sim-0.065$ dex per kpc) in the inner $\sim40$ arcsec; for KK189, within the inner $\sim30$ arcsec, the linear fit gives $\sim-0.36$ dex per arcmin ($\sim-0.3$ dex per kpc). Due to the large errorbars reported in Fig. \ref{mgrad}, we may conclude that an \emph{overall} metallicity gradient is definitely present for ESO269-66, while KKS57 and KK189 show hints of a gradient only in their central regions. Finally, for ESO269-66 we have no information about the stellar population of its nucleus, because it is not resolved in the photometry.

We may further want to check if there are differences in the stellar population of the galaxies, as observed in some early--type dwarfs in the Local Group. For each galaxy we thus divide the stars with derived metallicities in two subsamples. The first (\emph{metal--poor}) subsample contains stars with metallicity values $<(<$[Fe/H]$>_{med}-0.2)$, while the second (\emph{metal--rich}) subsample has values $>(<$[Fe/H]$>_{med}+0.2)$. We thus avoid any significant overlap for the subsamples, by excluding the values around the peak of the metallicity distribution functions.

First of all, we check the results derived for the metallicity gradients by plotting the ratio of metal--rich to metal--poor stars as a function of elliptical radius (Fig. \ref{mrtomp}, where the errorbars come from the Monte Carlo realizations described above). The ratio is overall decreasing with radius for all of the galaxies. For KKS57, CenN and KK189, this is valid out to a bit more than the $r_{h}$, after which the ratio tends to be dominated by statistical errors and fluctuations. These are due to the small-number statistics and to the fact that we are averaging over an elliptical radius, thus not taking into account asymmetric features. For the other three galaxies, the ratio decreases over their whole extent.

\subsection{Population gradients?}

\begin{figure*}
 \centering
  \includegraphics[width=7.5cm]{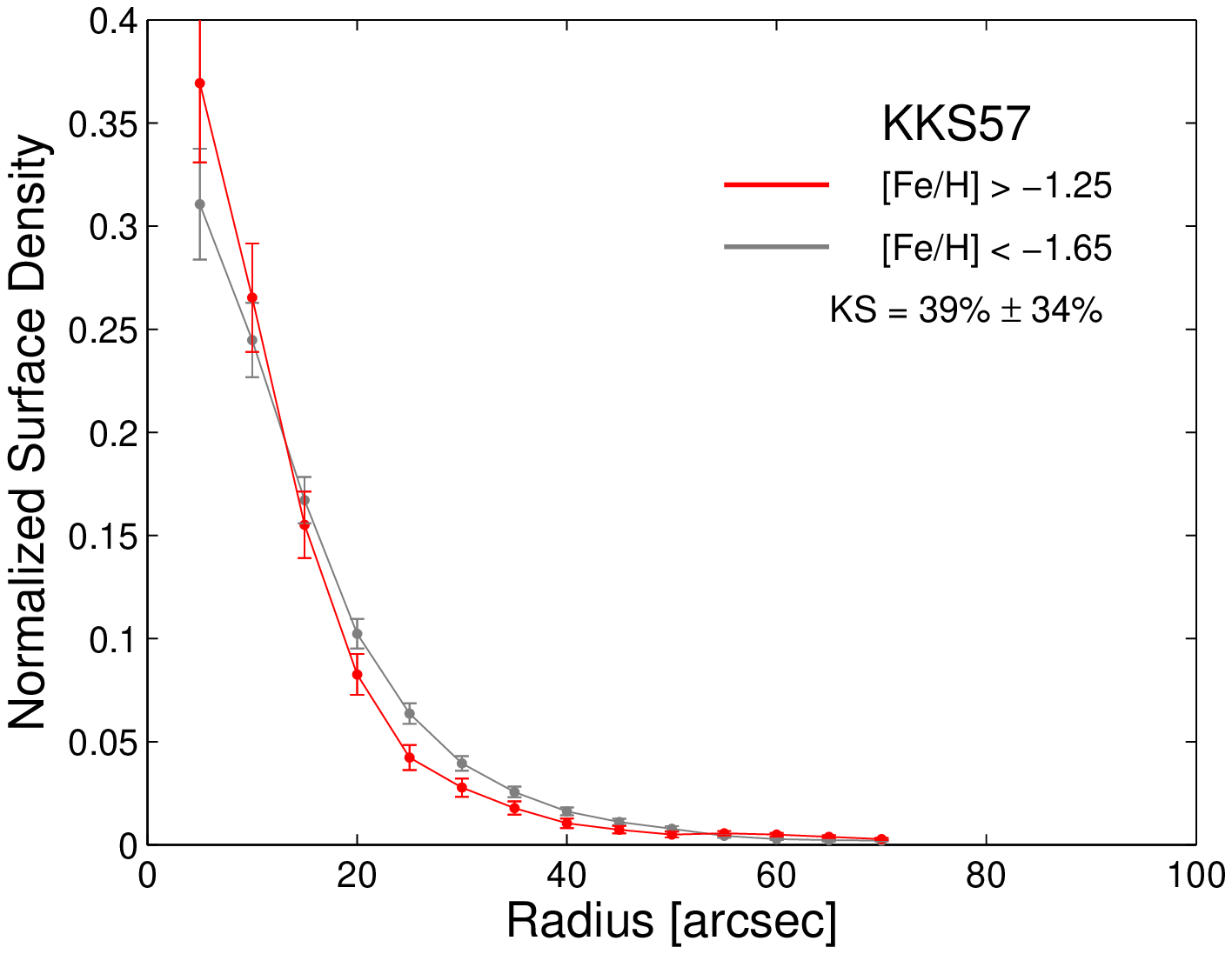}
  \includegraphics[width=7.5cm]{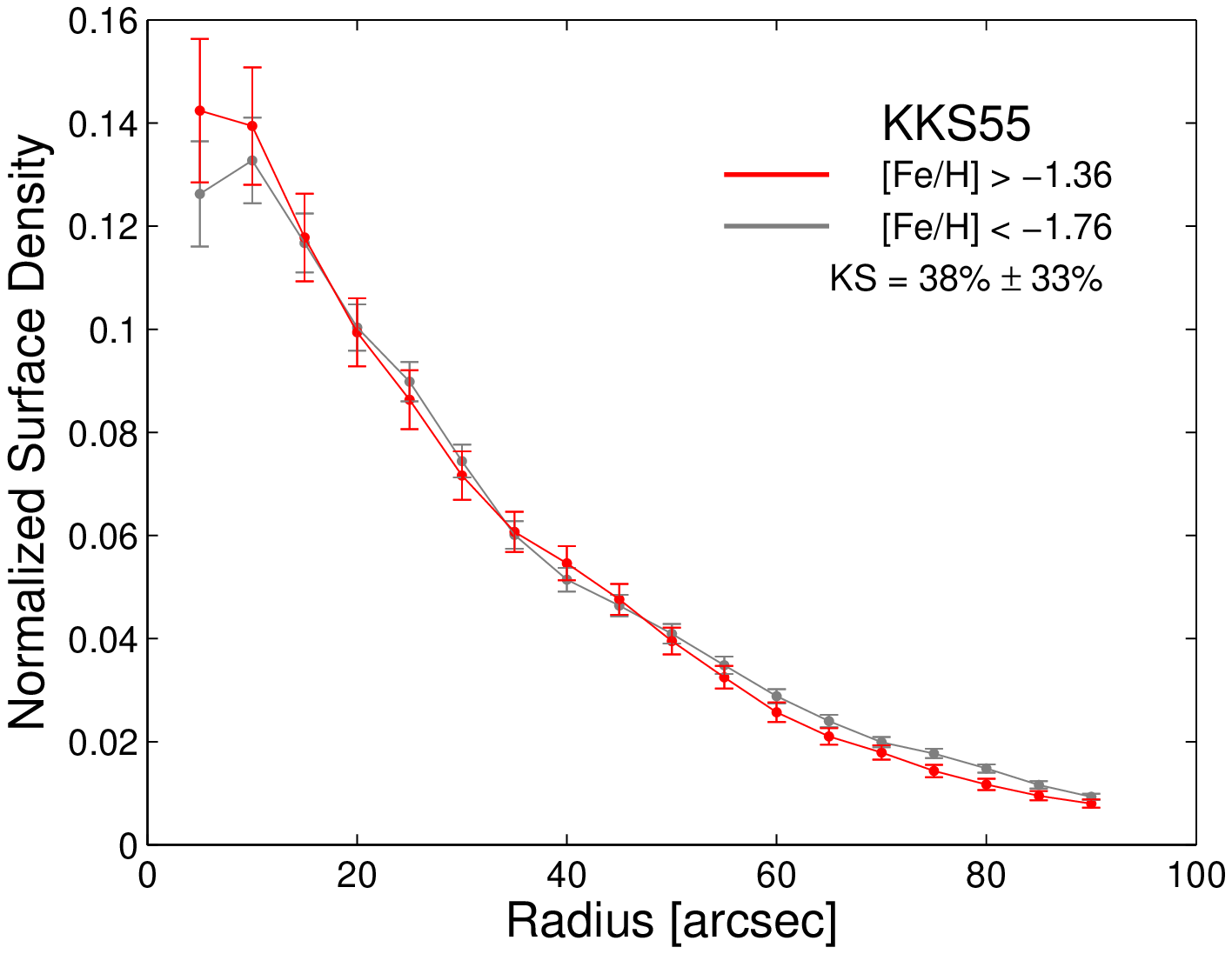}
  \includegraphics[width=7.5cm]{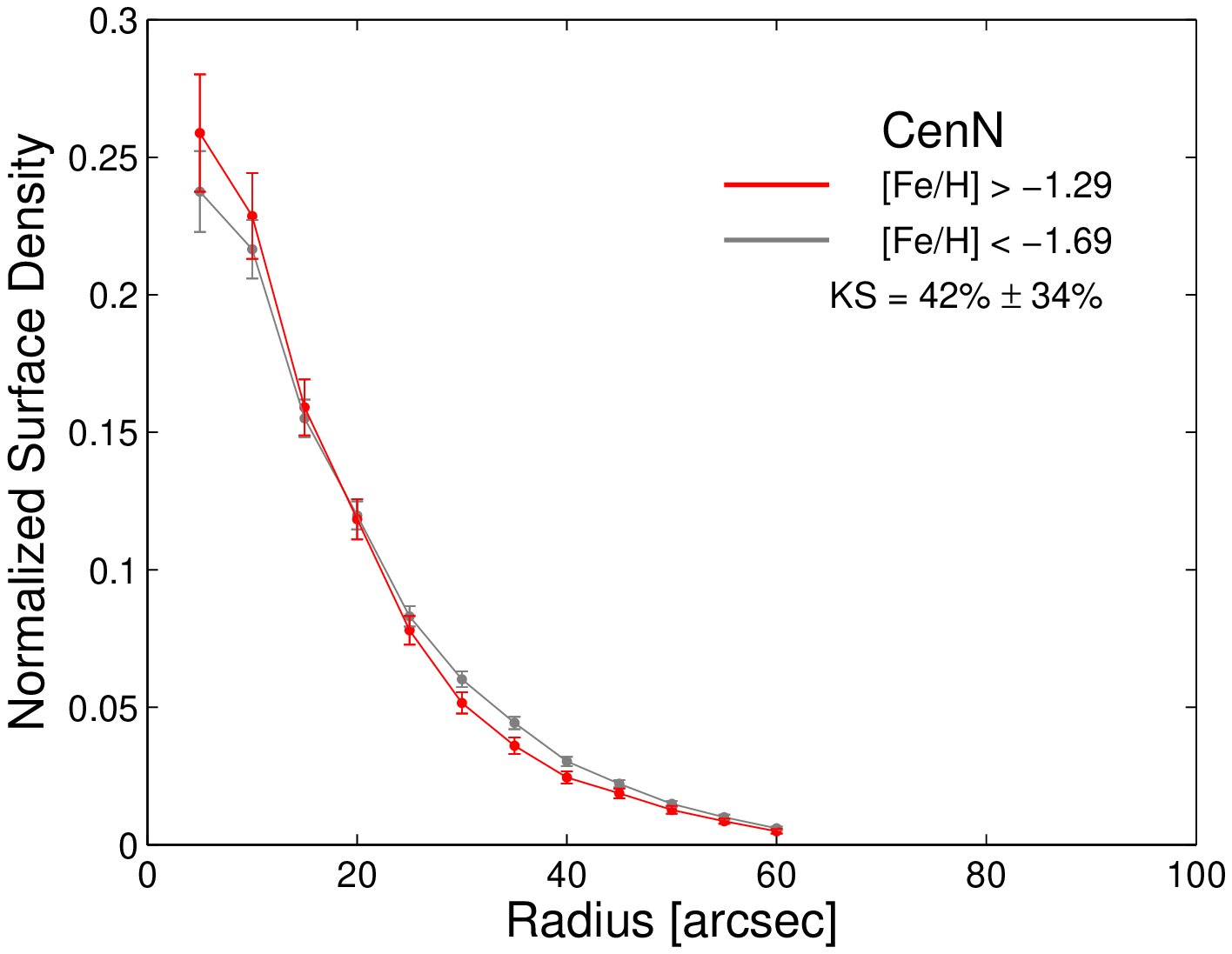}
  \includegraphics[width=7.5cm]{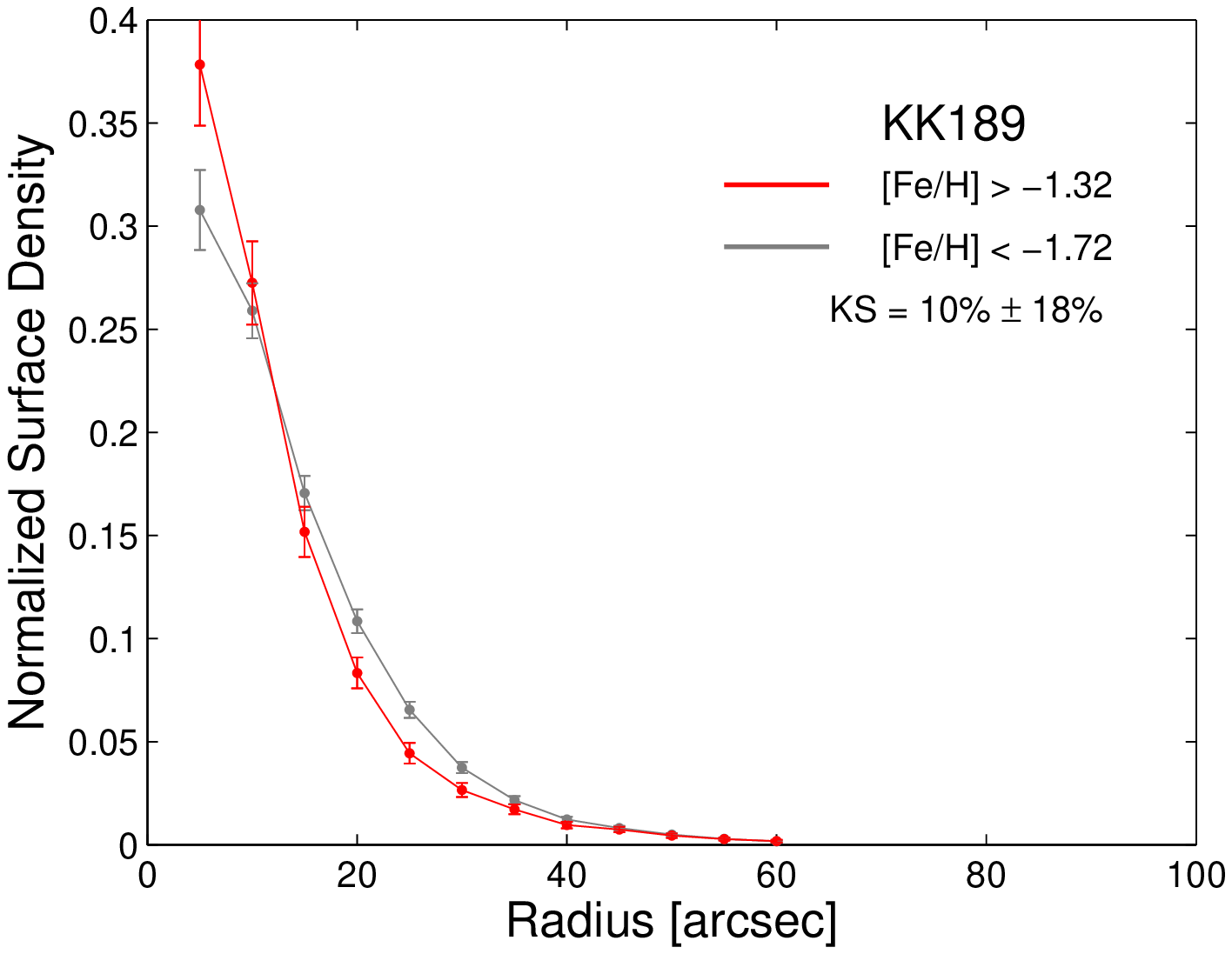}
  \includegraphics[width=7.5cm]{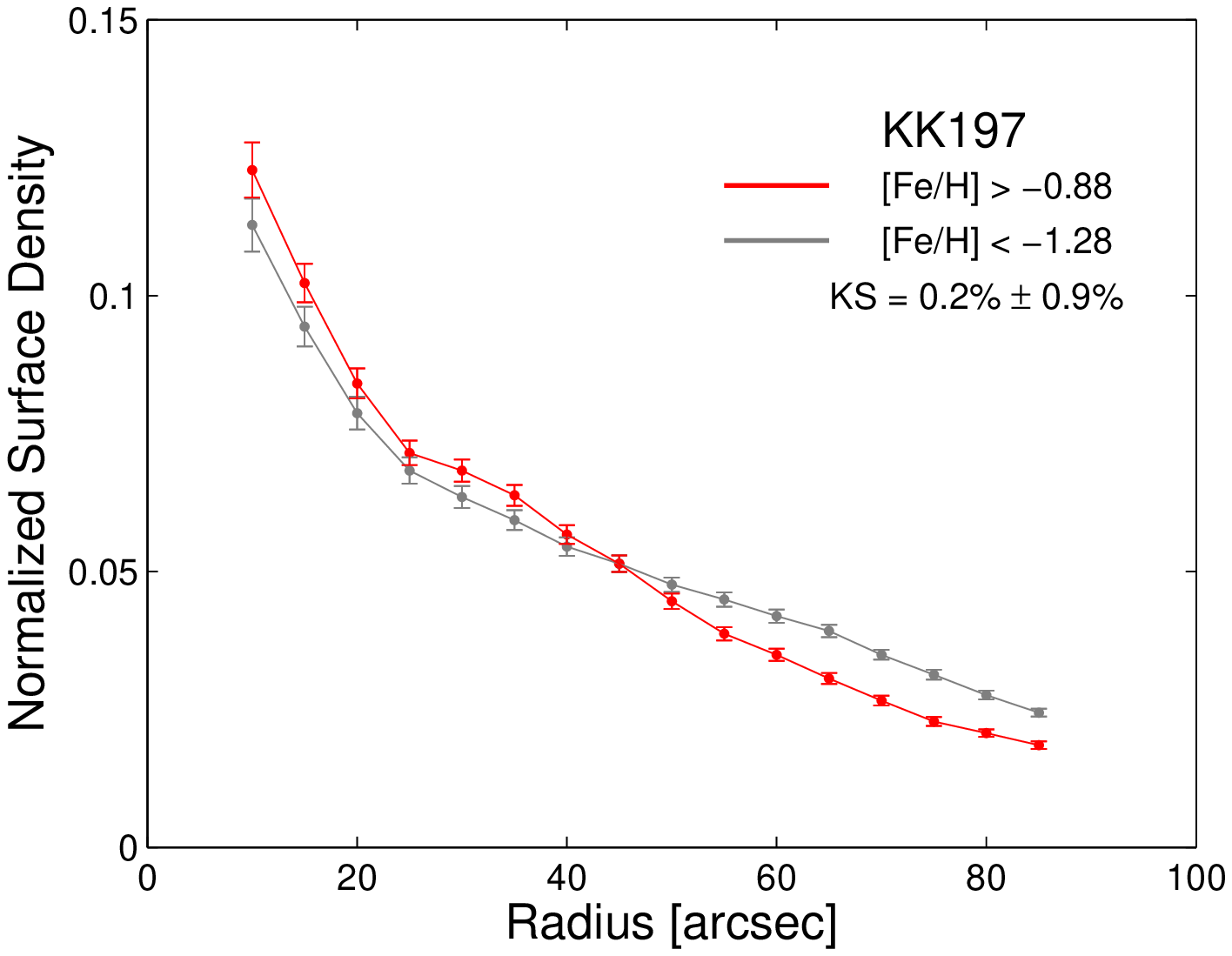}
  \includegraphics[width=7.5cm]{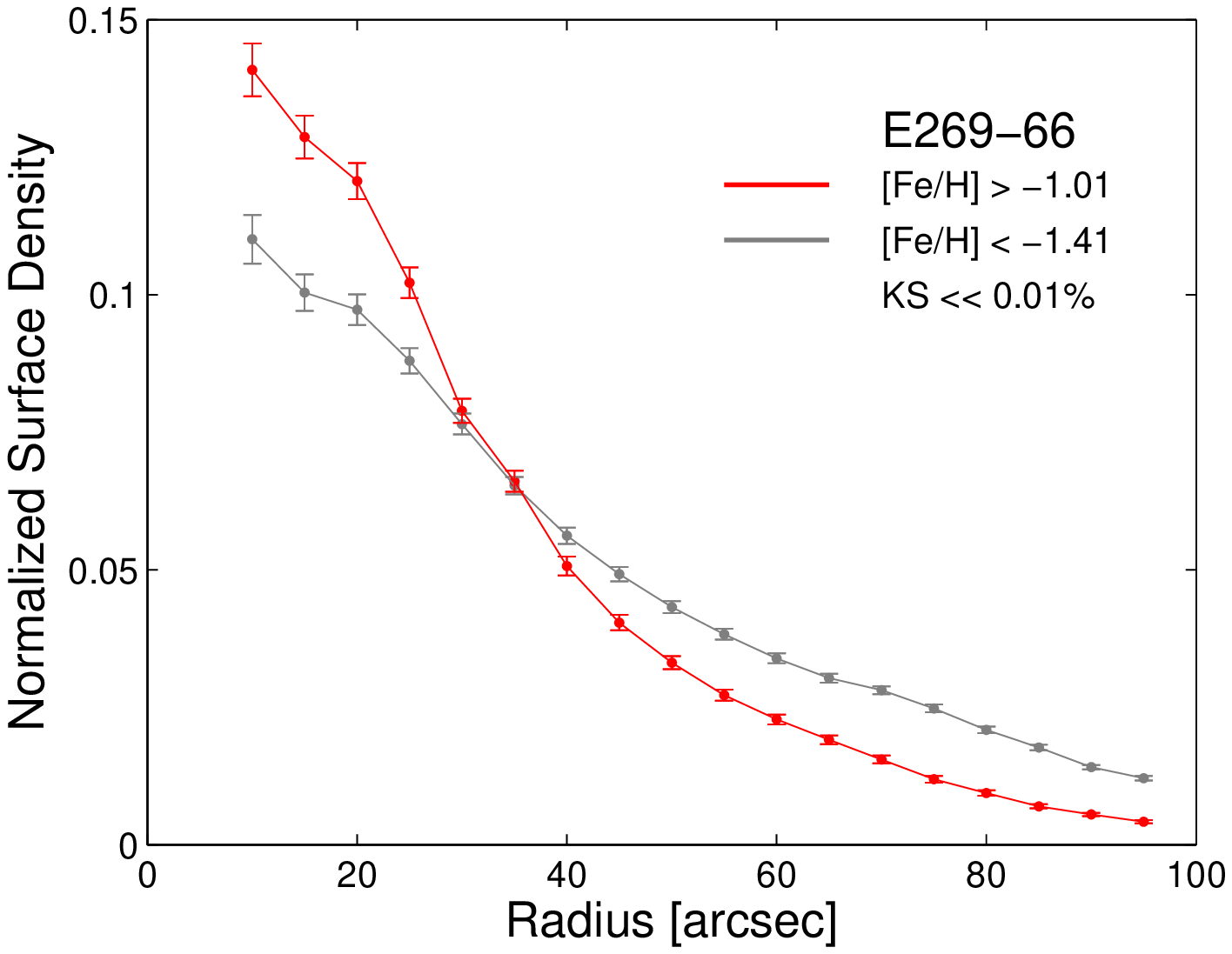}
 \caption{\footnotesize{Normalized surface brightness profiles as a function of elliptical radius. For each galaxy there is a distinction between \emph{metal--poor} (solid grey lines) and \emph{metal--rich} (solid red lines) stars (with metallicities values $<(<$[Fe/H]$>_{med}-0.2)$ and $>(<$[Fe/H]$>_{med}+0.2)$, respectively). We also report the result of a KS test on the cumulative distribution functions of the two subsamples, in order to statistically assess whether they belong to the same parent population or not (see text for details).}}
 \label{sbps}
\end{figure*}

\begin{figure*}
 \centering
  {\includegraphics[width=8cm]{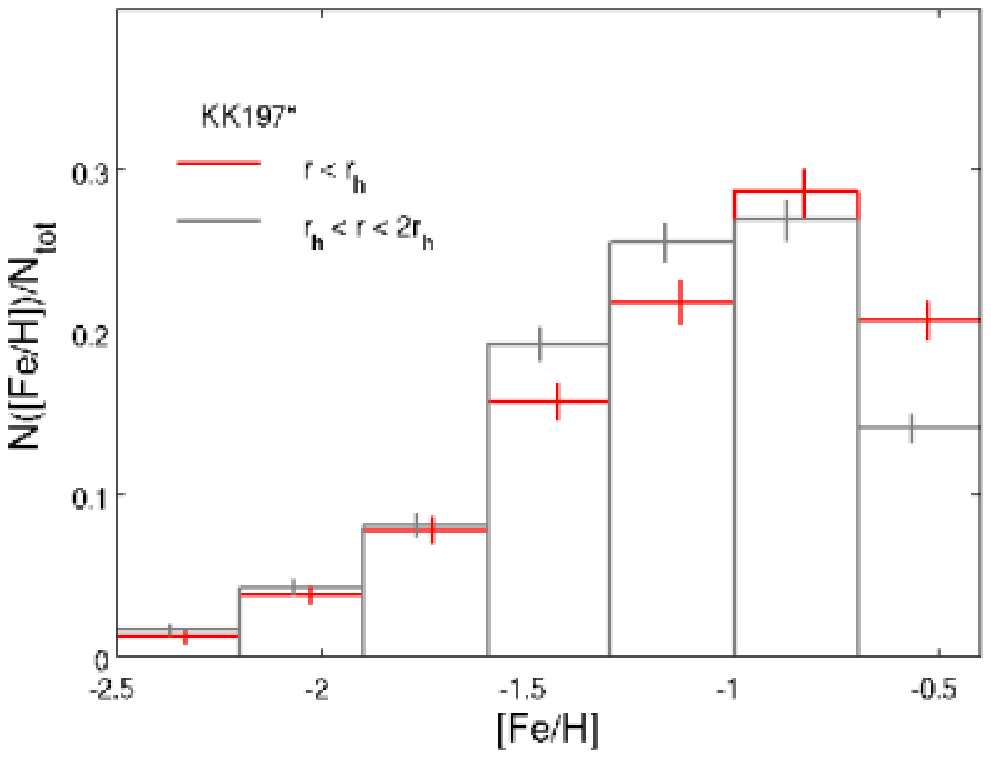}
  \includegraphics[width=8cm]{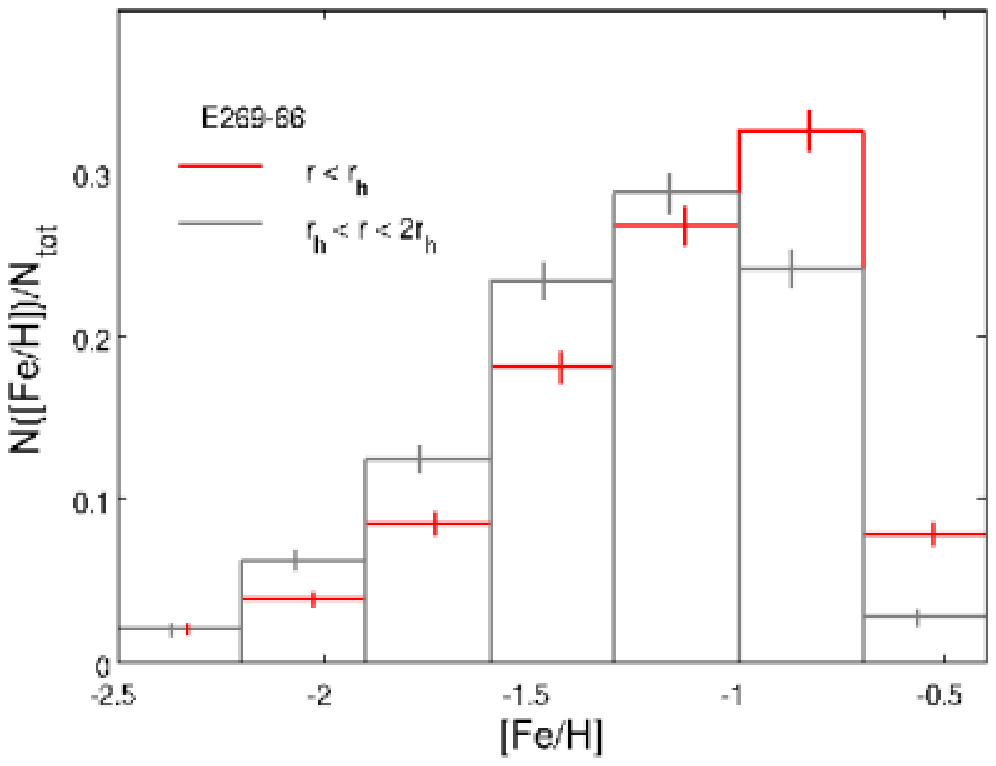}}
 \caption{\footnotesize{Normalized metallicity distribution functions for the inner and outer regions of KK197 and E269-66. The RGB stars are divided in two subsamples, respectively with distances of $r<r_{h}$ and $r_{h}<r<2r_{h}$ from the galaxy center. For KK197 the data do not go out to $2r_{h}$, so the second subsample contains stars with distances that go from $r>r_{h}$ until the borders of the CDD frame (corresponding to $\sim1.35r_{h}$).}}
 \label{mdf2}
\end{figure*}

\begin{figure*}
 \centering
  \includegraphics[width=7.5cm]{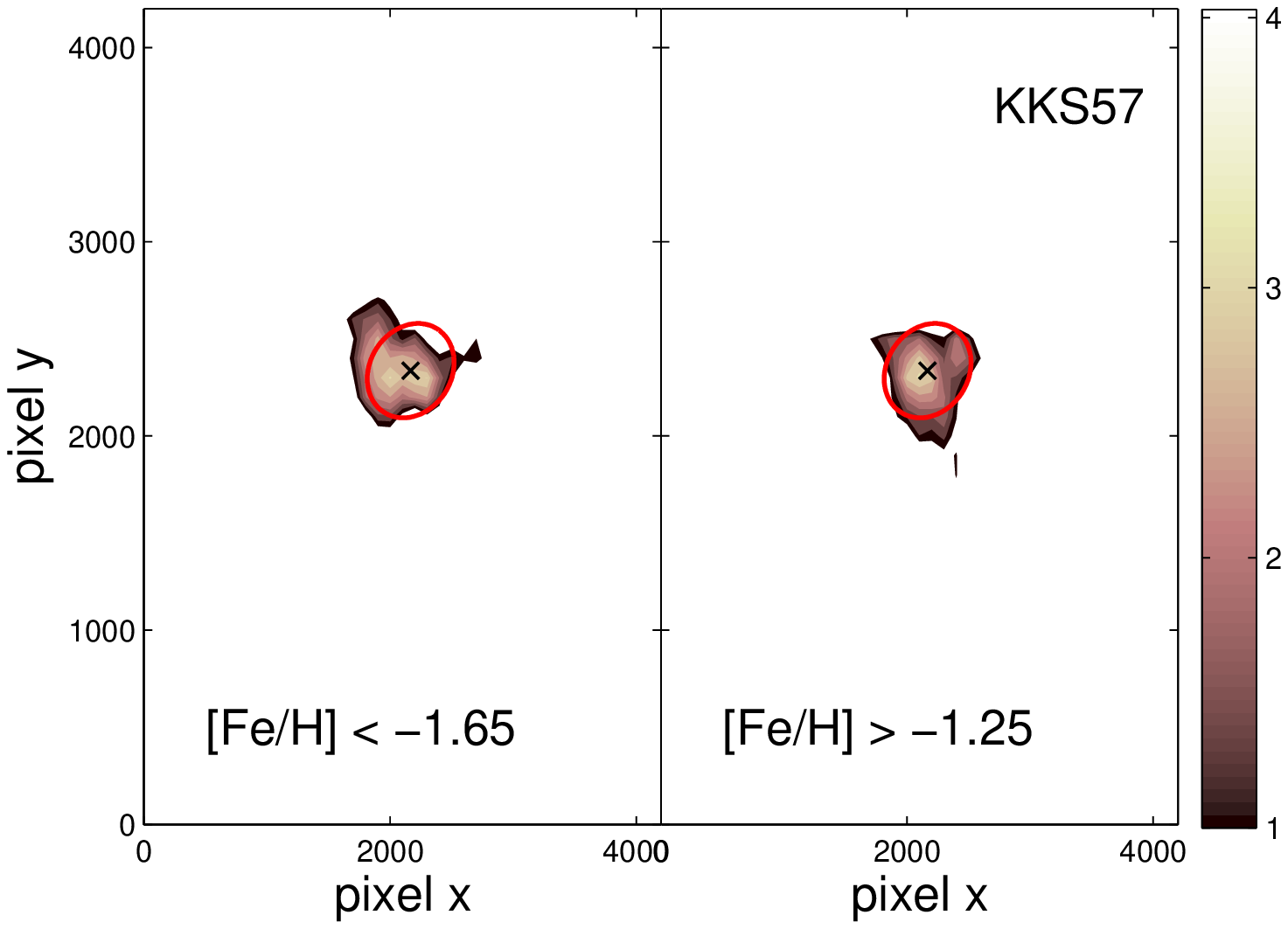}
  \includegraphics[width=7.5cm]{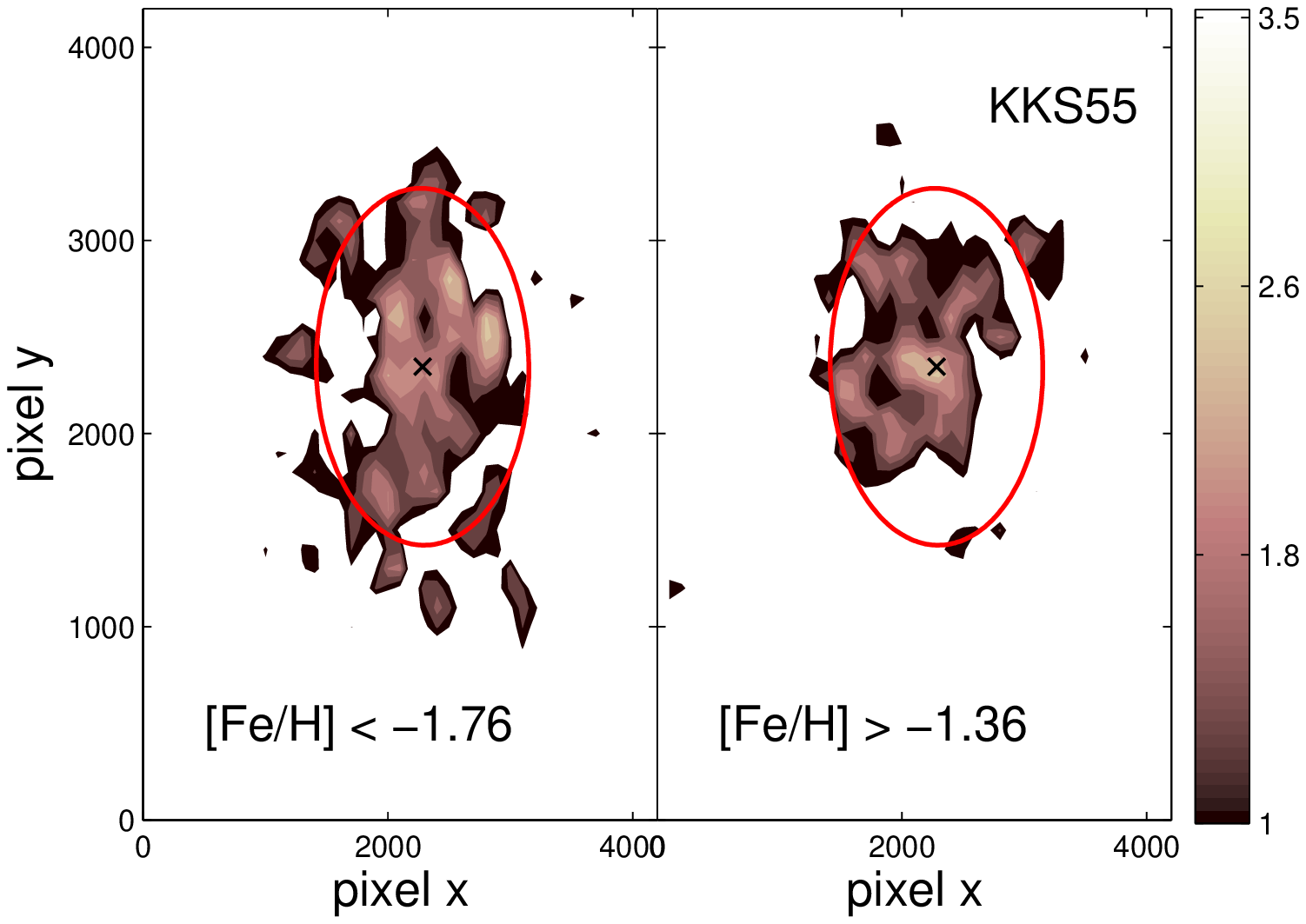}
  \includegraphics[width=7.5cm]{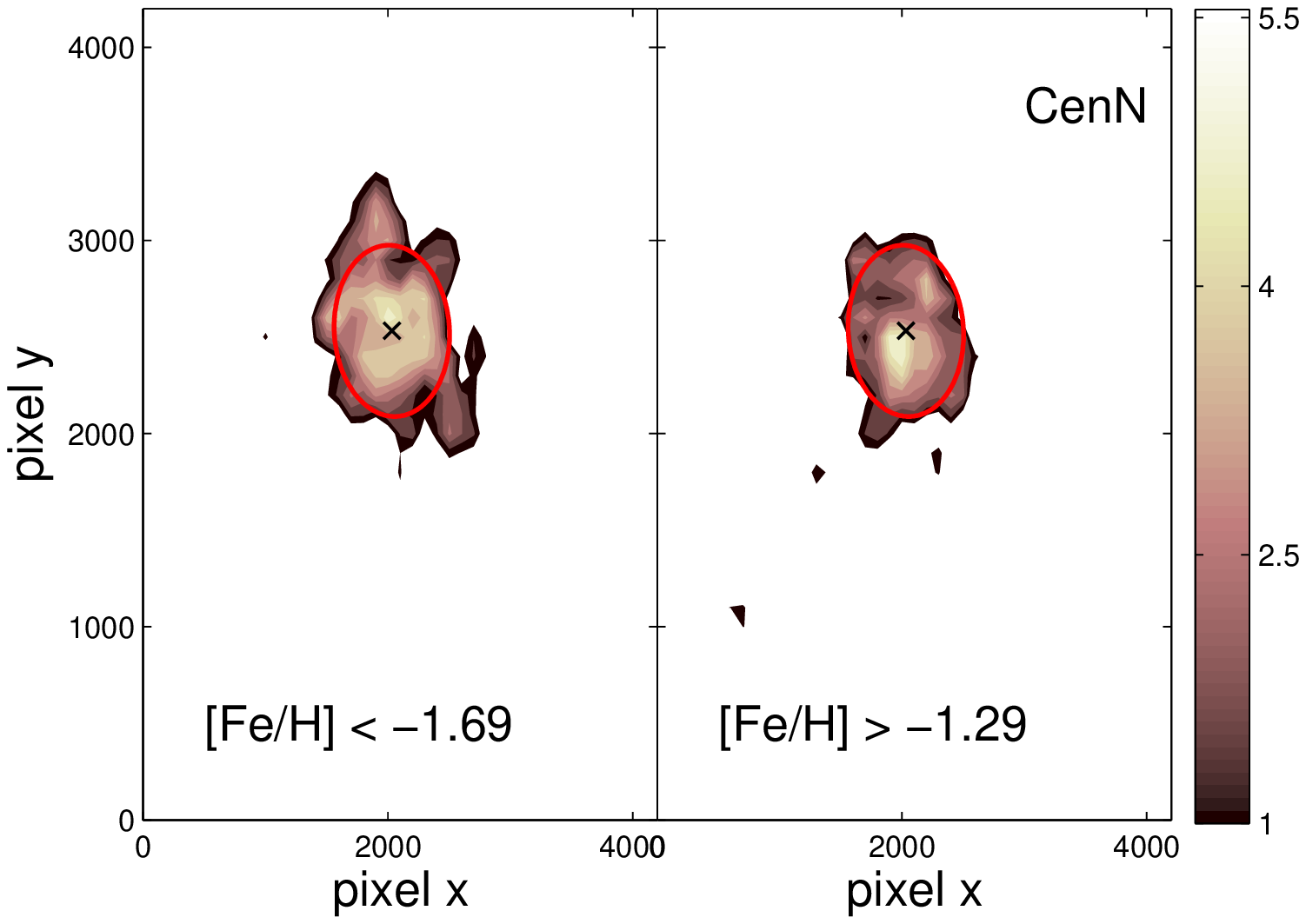}
  \includegraphics[width=7.5cm]{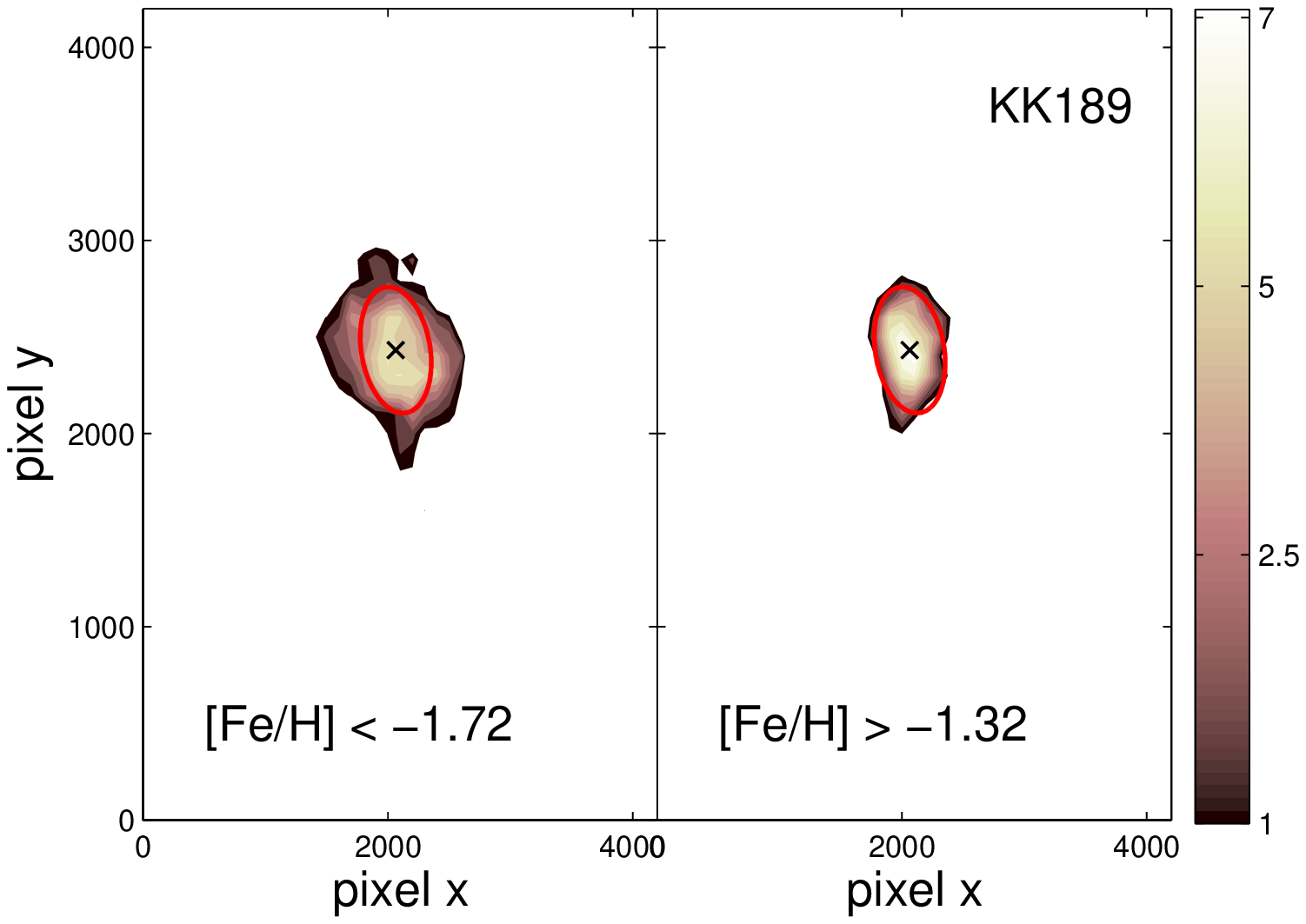}
  \includegraphics[width=7.5cm]{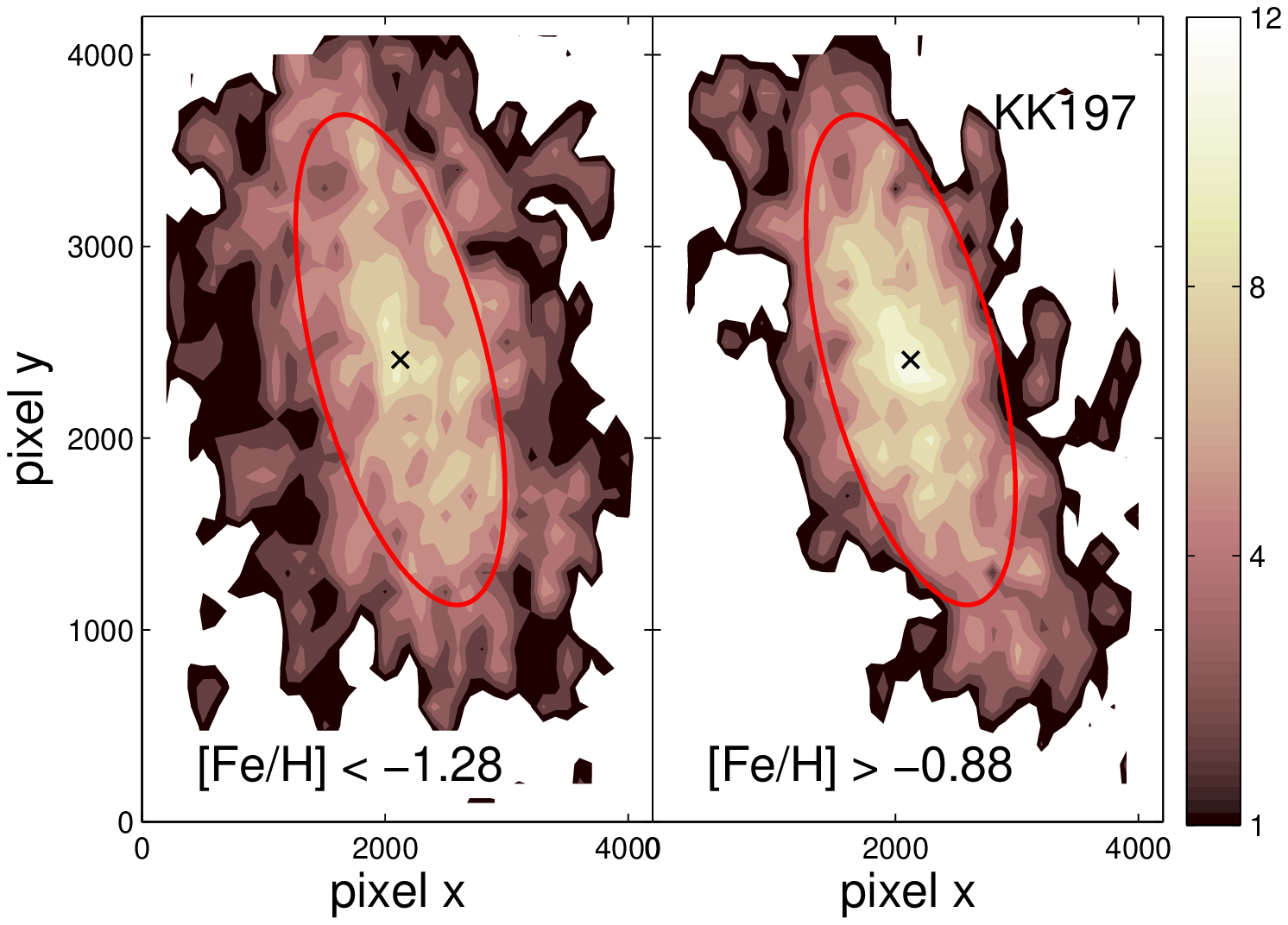}
  \includegraphics[width=7.5cm]{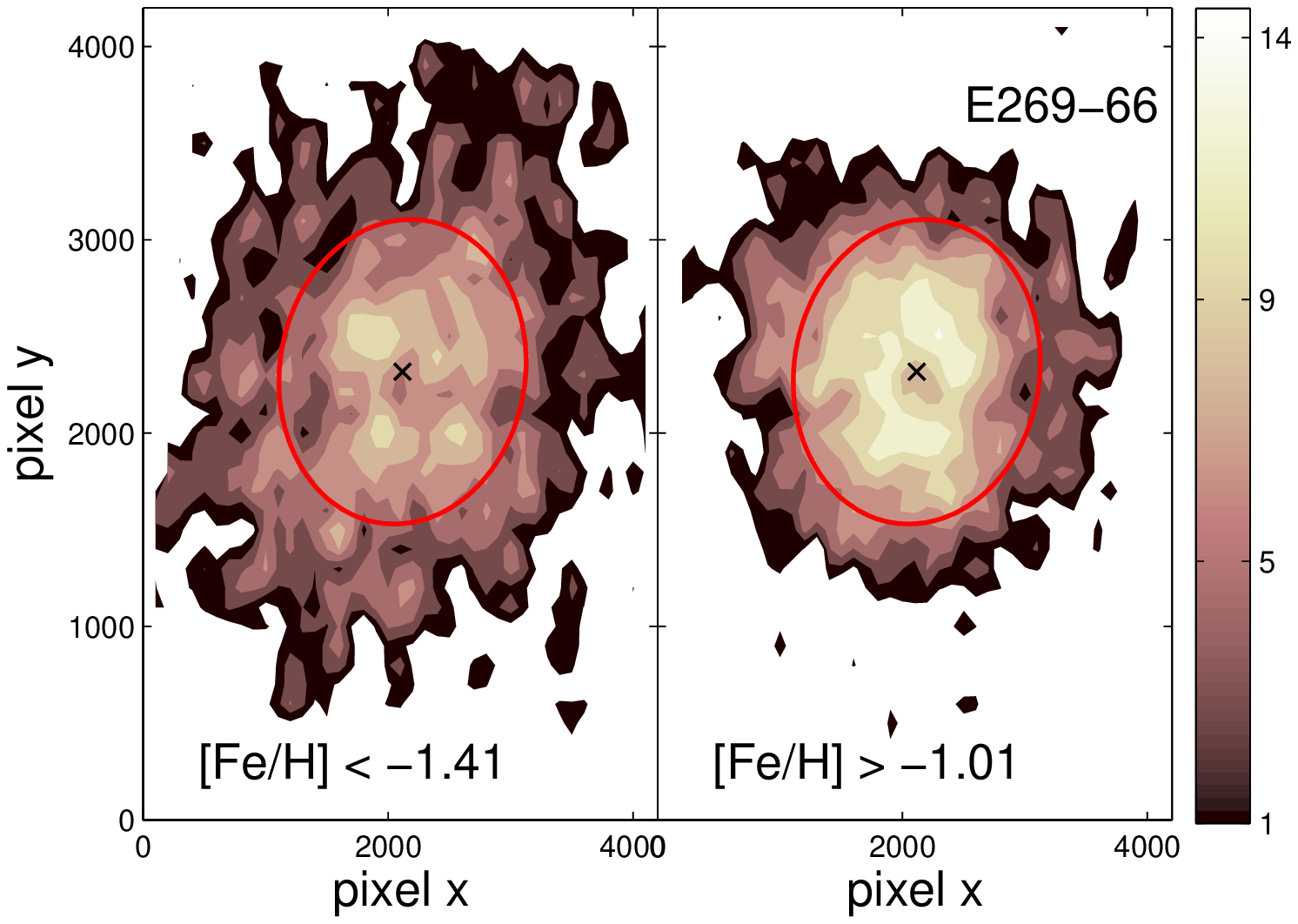}
 \caption{\footnotesize{Density maps for the galaxies, each diveded in two subsamples. For each galaxy, the left (right) panel shows \emph{metal--poor} (\emph{metal--rich}) stars with metallicity values $<(<$[Fe/H]$>_{med}-0.2)$ ($>(<$[Fe/H]$>_{med}+0.2)$). The color scale is the same for the two subsamples, normalized to the peak density value of the \emph{metal--rich} one, and the density values are listed in units of number of stars per $0.01$ kpc$^{2}$. 10 equally spaced isodensity contours are drawn starting at the $1\sigma$ up to the $3\sigma$ significance level ($4\sigma$ for KK197 and ESO269-66). The center of each galaxy is indicated with a black cross, and its half--light radius with a red ellipse (see Sect. \ref{ssd_sec} for details).}}
 \label{denmap}
\end{figure*}

For the two metallicity subsamples, we then derive $I$-band surface brightness profiles from star counts (corrected for radial incompleteness), and plot them in Fig. \ref{sbps} (normalized to the total number of considered stars). Indeed, for KK197 and ESO269-66 we clearly see a difference between the two subpopulations. To quantitatively test whether the two subpopulations are truly separated, we perform two-sided Kolmogorov--Smirnov (KS) tests, again varying the metallicity of each star within its metallicity error. In this way we can assess the robustness of our results. The values derived from the KS tests are reported along the surface brightness profiles in Fig. \ref{sbps}: for KKS57, CenN and KKS55, there is a probability of $\sim40\%$ (i.e., less than $1 \sigma$ significance level) that the two subsamples come from the same parent distribution; for KK189, this probability goes down to $\sim10\%$ ($\sim2 \sigma$); finally, for KK197 and ESO269-66 the null hypotesis can be rejected (with a probability of, respectively, $0.15\%$ and $<<0.01\%$), thus these galaxies do show significantly separated subpopulations. This is a robust result, even though our subdivision into subsamples is arbitrary and the radial distribution of the subsamples is partly overlapping, but with these data we have no way of investigating whether \emph{more} than two stellar subpopulations are present.

\citet{jerjen00b} performed surface photometry in the $B$ and $R$-bands of dwarfs in the Centaurus A and Sculptor groups. Their sample contains three of our galaxies (KK189, KK197 and ESO269-66). \citet{jerjen00b} had a slightly larger field of view for their observations than provided by our ACS data, but the results are consistent with ours. It is particularly interesting to look at their $B-R$ color profiles as a function of radius: for KK189 the color profile stays constant, while for the other two dwarfs in common with our study they become redder for radial distances from the center greater than the projected half--light radius. This supports our results regarding an older/more metal--poor population that dominates the outskirts of these galaxies.

We can also divide the stars in two subsamples depending on their distance from the galaxy center, e.g., $r<r_{h}$ and $r_{h}<r<2r_{h}$. This is done in order to compare the results for each galaxy with a fixed physical quantity, the half--light radius, and in order to be able to do such comparisons in the future for other studies. We chose the ranges such that there is a sufficiently large number of stars in each sample. The ACS data cover (at least) $2r_{h}$ for (almost) all of our target galaxies. The only exception is KK197, for which the data cover less than $2r_{h}$, so that the stars in the second subsample have distances ranging from $r_{h}$ up to $\sim1.35r_{h}$. We check for the less extended galaxies whether the results change significantly when choosing a larger second interval ($r_{h}<r<3r_{h}$), but we do not see evidence for that. We then derive normalized (to the total number of considered stars) MDFs for the two radius-selected subsamples. For the galaxies for which there is no significant signature of two different subpopulations, the normalized MDFs are indistinguishable within the errorbars, but for KK197 and ESO269-66, they show a marked difference as expected. We report the normalized MDFs for the two latter dwarfs in Fig. \ref{mdf2}. The fact that the data do not cover the whole extent of KK197 within $2r_{h}$ does not change our conclusions, since a gradient is already clearly present in the inner regions of the galaxy. KK197 and ESO269-66 are classified as dwarf ellipticals, they are the most luminous galaxies in our sample, and they are located very close to the central dominant galaxy of the group (see Fig. \ref{sky}).

However, in these kinds of analyses one has the keep in mind two aspects. Firstly, by looking at properties as a function of projected radius, possible non--radial gradients can be averaged out. We thus also compute area density maps for the two subsamples in the following way. At the mean distance of the group ($\sim3.8$ Mpc), 1 arcsec approximately corresponds to $\sim0.02$ kpc. We first count the number of neighbours within $\sim0.03$ kpc$^{2}$. This value was chosen in order to avoid introducing any small-scale noise or artificial substructures, but to permit us to still recognize overall features. Then we convolve the results with a square grid and get a final resolution of $0.01$ kpc$^{2}$. The resulting density maps are shown in Fig. \ref{denmap}. The color scale is the same for both subsamples of each galaxy, normalized to the peak density value of the \emph{metal--rich} sample, and the values are reported in units of number of stars per $0.01$ kpc$^{2}$. We draw 10 equally spaced isodensity contour levels starting at the $1\sigma$ significance level up to the $3\sigma$ significance level ($4\sigma$ for KK197 and ESO269-66). In every case, the outer contours have values of many ($\sim5$ on average) $\sigma$ above the Galactic foreground level.

The density maps of the two subsamples in each galaxy do not show striking differences. In every case, the subsamples' distributions follow the elongation of the galaxy. However, we can see that the \emph{metal--rich} stars tend to be more centrally concentrated than the \emph{metal--poor} ones in KK197 and ESO269-66. There are some fluctuations and asymmetries within the galaxies, to some extent due to small-scale substructure and poor sampling in the outskirts. In the most massive galaxies, the ACS gap between its two CCDs is visible in the density maps (a horizontal line at pixel y $\sim2000$), but this does not affect their general features. For KK197, an unresolved globular cluster is found near the position of the central peak \citep{georgiev08}, which may in fact represent a nucleus. For the nucleated galaxy ESO269-66, a depression in the density map is seen at the center, because we chose to leave out the nucleus due to the high crowding in that region.

\begin{figure*}
 \centering
  \includegraphics[width=14cm]{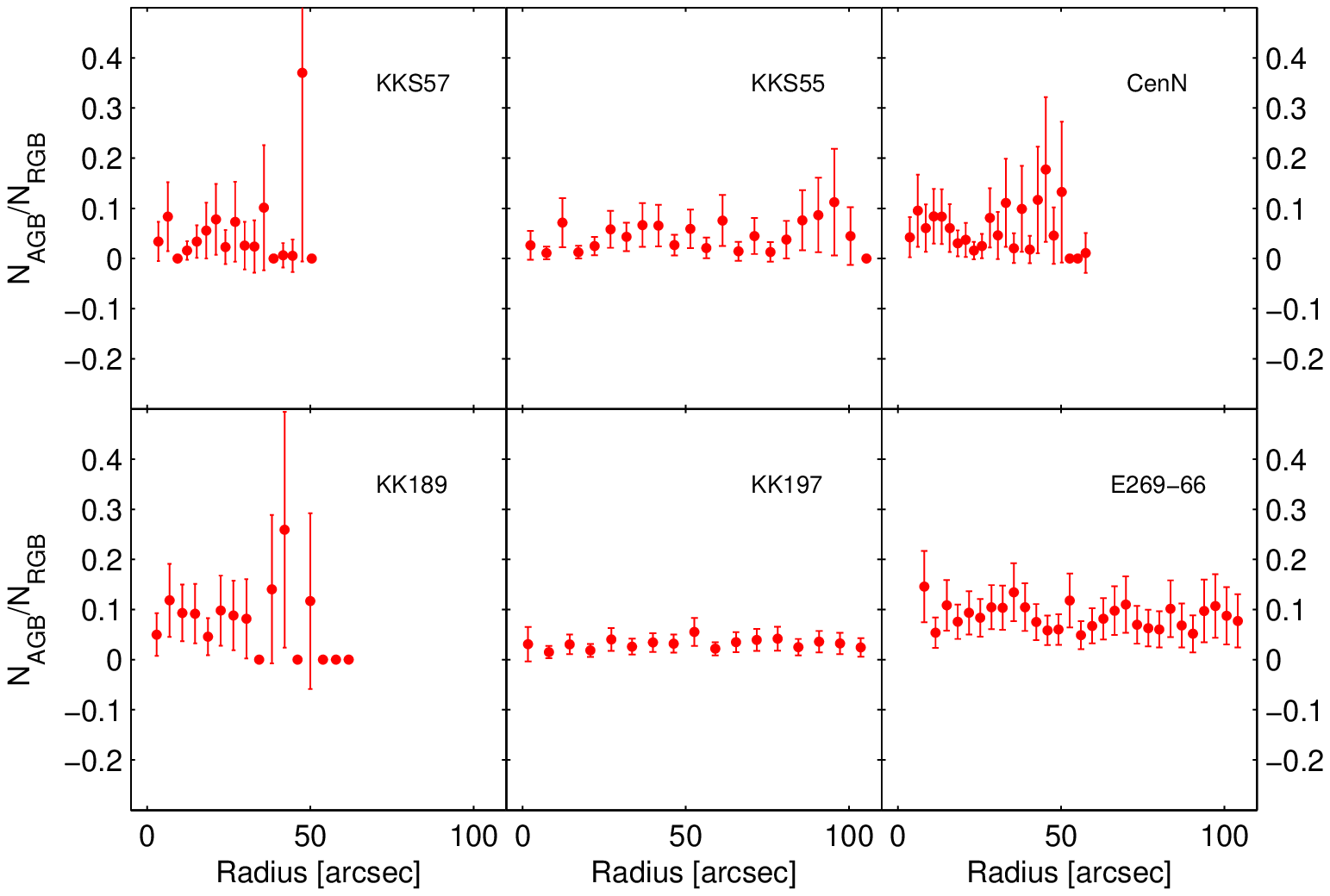}
 \caption{\footnotesize{Ratio of luminous AGB to RGB stars as a function of elliptical radius (see text for details about the selection of the two subsamples).}}
 \label{agbtorgb}
\end{figure*}

The second thing that has to be mentioned is that, as seen in the previous Sect., for some of our galaxies there is a non negligible presence of an intermediate--age population. The intermediate--age stars could, in principle, bias the derived metallicity gradients. We thus have a qualitative look at the radial distribution of the luminous AGB stars. We compute the number of AGB and RGB stars (selected as described in Sec. \ref{mdf_sec}) in elliptical annuli of $\sim4$ arcsec width. We correct the counts for radial incompleteness, and we also subtract the estimated number of contaminants for AGB and RGB separately. This has to be done because the amount of relative contamination is quite different for the two samples due to the smaller number of luminous AGB stars with respect the RGB stars. We then compute the ratio between AGB and RGB stars, and consider it as a function of radius. The results are plotted in Fig. \ref{agbtorgb}. The strong fluctuations in these plots are just reflecting the poor statistics. In almost every case, this ratio stays nearly constant over the whole extent of the galaxy. Only for ESO269-66, the inner $\sim20$ arcsec show a slightly higher AGB to RGB ratio, revealing the presence of a small amount of intermediate--age AGB stars concentrated in the inner region of this galaxy. As discussed before, this younger population would - at least qualitatively - increase the number of apparently blue, metal--poor RGB stars, thus biasing the metal--poor subsample. Nonetheless, we see that the more metal--rich stars are clearly more centrally concentrated (Fig. \ref{sbps} and \ref{mdf2}). On the other side, the metallicity gradient that we find for this galaxy is still significant: it could be even more pronounced if some of the central stars that we identify as metal--poor RGB stars were in fact more metal--rich, low-luminosity AGB stars superimposed on the RGB.


\section{Discussion} \label{discuss}

\begin{figure}
 \centering
  {\includegraphics[width=9cm]{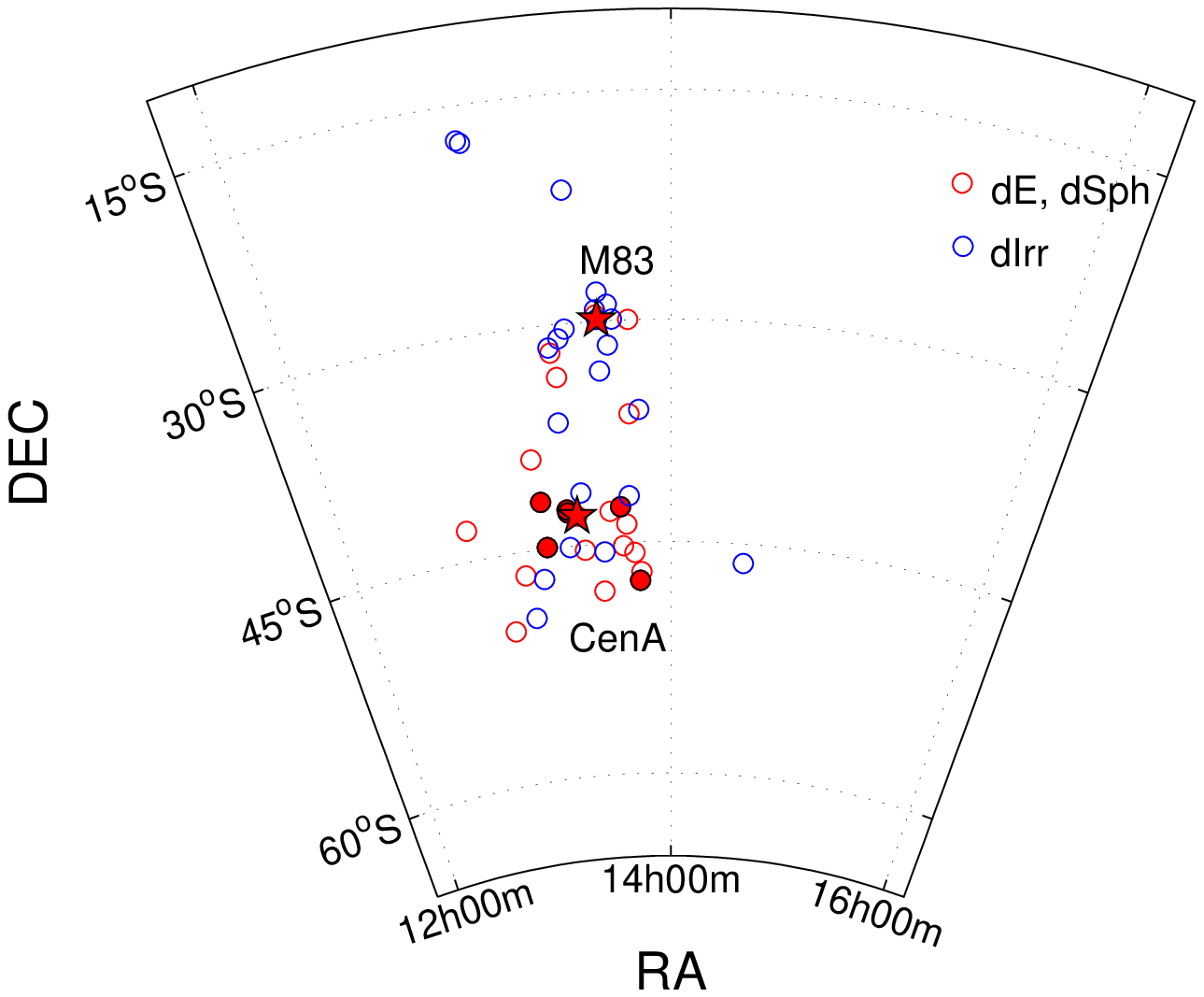}
  \includegraphics[width=9cm]{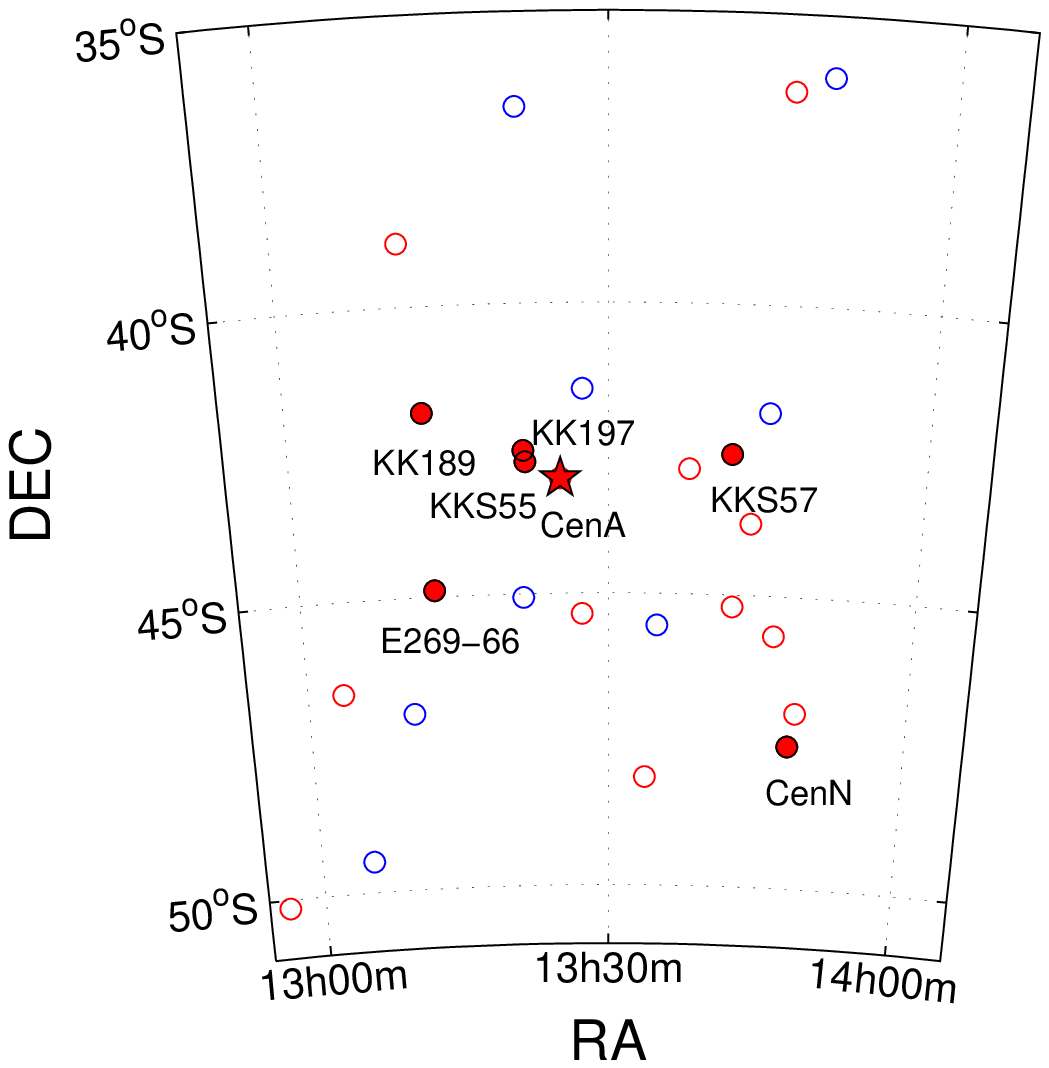}}
 \caption{\footnotesize{\emph{Upper panel.} Positions in the sky of the galaxies belonging to the Centaurus A group (from \citealt{kara07}). Red symbols indicate for early--type dwarfs (dE and dSph), while blue symbols are for late--type dwarfs (dIrr). Two red stars are drawn at the positions of the two dominant giant galaxies Centaurus A and M83, around which the smaller companions are clustering and forming two distinct subgroups. The filled circles represent the dwarfs studied here. \emph{Lower panel.} Same as above, just zoomed in in a smaller region around Centaurus A, where the galaxies studied here are located (as labeled in the plot).}}
 \label{sky}
\end{figure}

In order to illustrate where the target galaxies are located within the Centaurus A group, we plot the position in the sky of the galaxies in this group in Fig. \ref{sky}, using the data from \cite{kara07}. The upper panel shows the distribution of dwarfs with positive tidal index (i.e., belonging to the group) around the two giant galaxies Centaurus A and M83, which form two distinct subgroups. In the lower panel of the Figure, we zoom in around Centaurus A, since all of the galaxies studied here belong to the Centaurus A subgroup. This is obviously quite a dense region: it contains approximately the same number of galaxies within a fixed radius as our Local Group, with the difference that the census for Centaurus A is likely to be incomplete (e.g., \citealt{jerjen00b}, \citealt{rejkuba06}).

The six early--type dwarfs of our sample have a quite wide range of luminosities (see Table \ref{infogen}), and their morphological types vary from dwarf spheroidals (KKs57, KKs55, CenN and KK189), to dwarf ellipticals (KK197 and ESO269-66, the latter being a nucleated galaxy). What they have in common is the absence of any recent star formation (as traced by H$\alpha$, or young populations from CMD), nor is there any significant amount of HI detected that would allow for ongoing or future star formation. For example, \citet{bouchard07} investigated the gas content in three of our target dwarfs (KK189, KK197 and ESO269-66) and found only upper limits for their HI masses ($<1.6$, $<2.9$ and $<1.0\times 10^{5}$M$_{\odot}$, respectively). Moreover, only upper limits can be found for the star formation derived from H$\alpha$ emission: for KK189, KK197 and ESO269-66 the values reported by \citet{bouchard08} are $<0.4\times10^{-5}$M$_{\odot}$yr$^{-1}$.

The metallicities derived here can be compared to the work of \citet{sharina08}. These authors derive the mean metallicities for a large sample of nearby dwarfs, among which there are also our target objects. \citet{sharina08} apply the empirical formula [Fe/H]$=-12.64+12.6(V-I)_{-3.5}-3.3(V-I)^{2}_{-3.5}$ \citep{lee93}, which infers the mean metallicity from the mean color of the RGB at an absolute magnitude of M$_I=-3.5$. Their derived values are all slightly lower ($\sim10-15\%$ in dex) than the values found with our isochrone interpolation method, but still consistent within the errors, if we consider the difference ($\sim0.1$ dex) between the empirical metallicity scale from the above mentioned formula and the one used in the Dartmouth isochrone models. This discrepancy is also possibly due to a different selection of stars for the computation of the mean color of the RGB.

\citet{sharina08} further plot the color spread as a function of luminosity, and suggest that the linear correlation between these two quantities could be due to a more intense star formation in more massive galaxies, probably because of denser gas reservoirs. They however note a very high \emph{color} spread for the galaxy KK197, twice of what could be expected for its luminosity. \citet{sharina08} suggest this to be the consequence of a possible strong mass loss. If we plot our derived \emph{metallicity} spreads as a function of luminosity, KK197 appears indeed to be slightly displaced from the linear relation shown by the other galaxies, and so does also KKS57. Looking at the position of these two galaxies within the group (Fig. \ref{sky}), and also their deprojected distance from Centaurus A (see below), they are located not far from the giant elliptical, but we have no way of adding more information about possible environmental effects because nothing can be said about the orbits that the dwarfs had in the past. However, \citet{bouchard07} point out that KK197 and ESO269-66 could both be influenced by the closeby radio lobes of Centaurus A, through which they may pass during their orbits and which may have contributed to the removal of the neutral gas content. In fact, ESO269-66 appears to have a particularly low HI gas mass to visual luminosity ratio, suggesting that some external agent could have contributed to the loss of its gas content. Although KKS55 is (in deprojection) located very close to Centaurus A as well, it does seem to contain a slightly bigger fraction of intermediate--age stars, as compared to KKS57, KK197 and ESO269-66. This could possibly imply that this galaxy may now be on its first close approach to Centaurus A.

\begin{figure}
 \centering
\resizebox{\hsize}{!}
{\includegraphics[width=9cm]{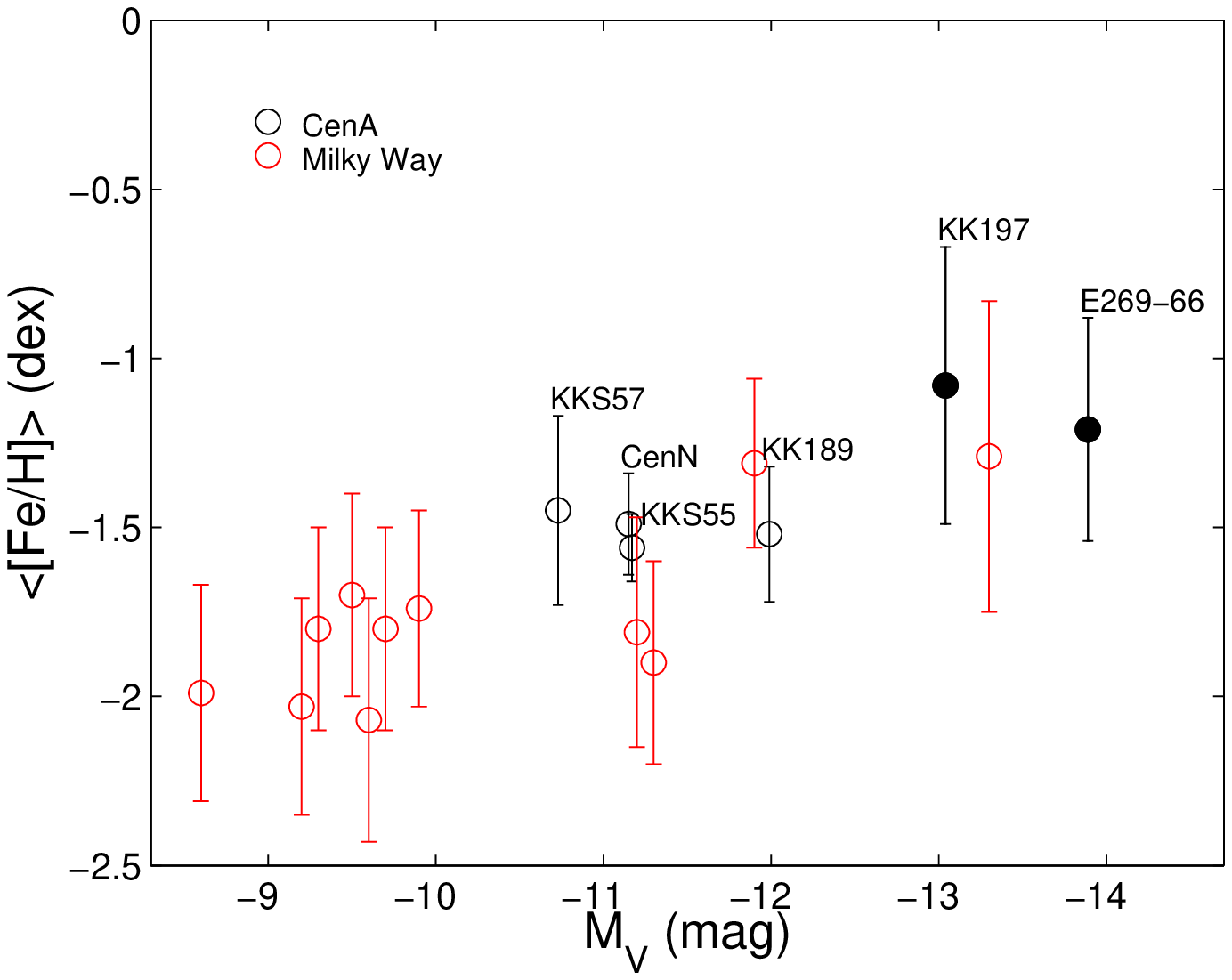}}
{\includegraphics[width=9cm]{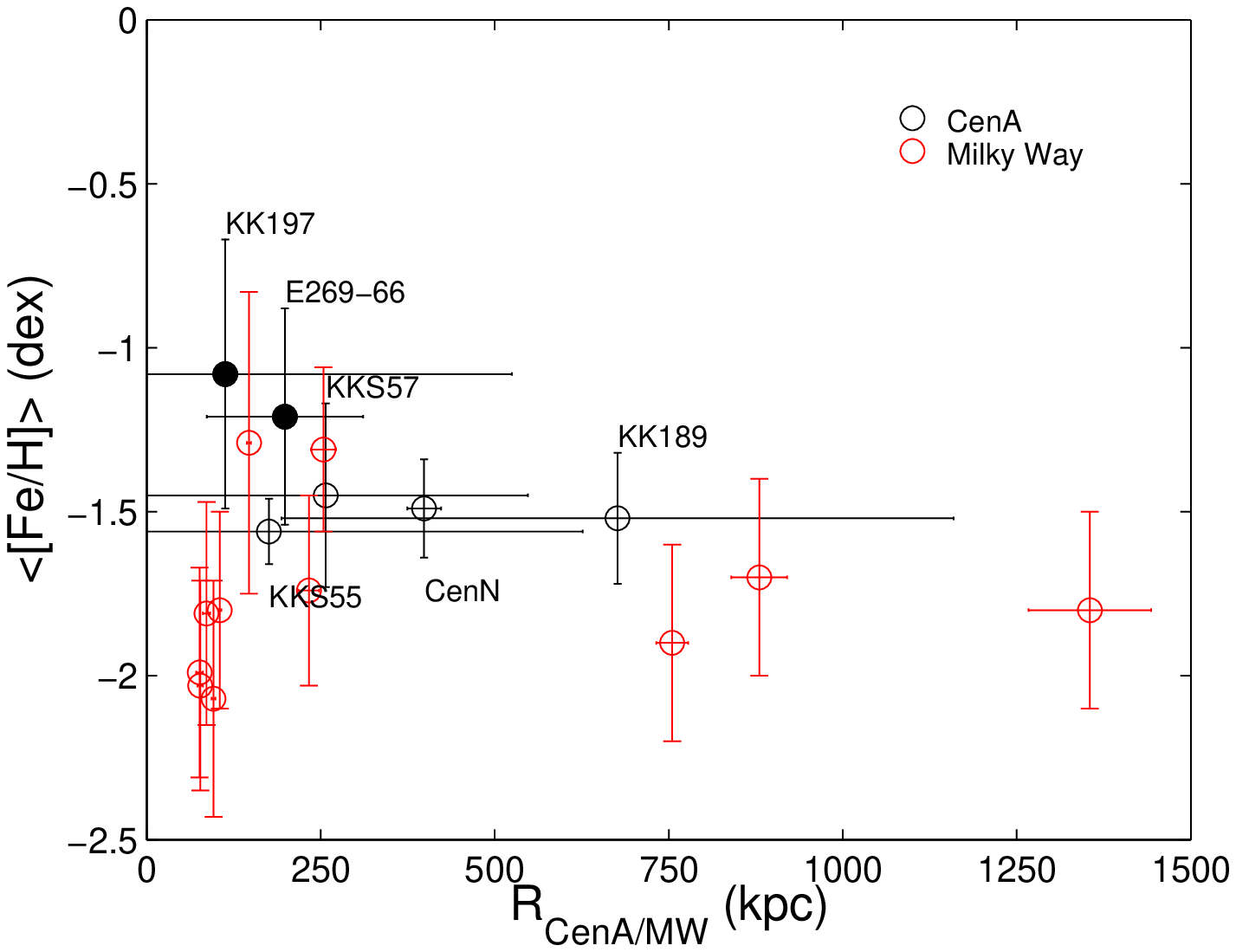}}
\caption{\footnotesize{Median metallicity plotted as a function of (\emph{upper panel}) absolute V luminosity and (\emph{lower panel}) deprojected distance from the dominant group galaxy. The black symbols are representing the Centaurus A group dwarfs studied here (filled circles are dwarf ellipticals, the others dwarf spheroidals), while red symbols are for Milky Way companions.}}
\label{fehvs}
\end{figure}

It is difficult, given the restricted amount of information we can derive from the data considered here, to make a detailed comparison to the dwarf galaxies of our own Local Group. However, there are some points that are suggesting that the smallest objects in these two groups may have had a similar history. Just as the complex and different star formation histories seen in our neighbourghood, we see that the metallicity and stellar population properties in our target dwarfs differ from each other (e.g., median metallicity, metallicity spread, fraction of intermediate--age stars). However, it looks like in our sample there are no purely old dwarfs such as Ursa Minor or Draco in the Local Group.

Coming to our results, all of the studied dwarfs are relatively metal--poor, as is expected for galaxies of such low luminosities. For comparison, dwarf spheroidals in the Local Group that lie in the same range of luminosities have mean values of [Fe/H]$=-1.7$ to $-1.2$ (see, e.g., \citealt{mateo98}; \citealt{grebel03}; \citealt{koch09}), thus being on average slightly more metal--poor than our sample (see Fig. \ref{fehvs}). This small difference could be due to the measurement techniques, from one side detailed spectroscopy and from the other side photometry. Moreover, as mentioned in Sect. \ref{mdf_sec}, the metallicities derived here could possibly be upper limits due to our assumptions about age and $\alpha$-enhancement, so we conclude that on average there are no significant differences between the dwarfs of the two groups. Dwarf galaxies with even lower luminosities have recently been found in the Local Group, which have even lower metallicities, but such objects are intrinsically too faint to be observed in the existing surveys at the Centaurus A group distance. This naturally leads us to the metallicity--luminosity relation, already analysed in detail for the Local Group and other environments by many authors (e.g. \citealt{mateo98}; \citealt{grebel03}; \citealt{thomas03}; \citealt{lee07}; \citealt{sharina08}). The six galaxies analysed here were already included in the metallicity--luminosity relation of \citet{sharina08}, their Fig. 9. They show that this relation is very similar for the dwarfs in the Local Group and those in other groups and in the field within $\sim10$ Mpc distance. They also comment on a few early--type dwarfs that show metallicities lower than all of the others, pointing out that they resemble dwarf irregulars in their metallicity content and could thus be in a transition phase. Among the galaxies with metallicity values lower than expected, there are the three galaxies in common with our study (KKs55, CenN and KK189). In the upper panel of Fig. \ref{fehvs}, we plot again the metallicity--luminosity relation with the small sample of values derived in our study (black circles) and adding the values for dwarf spheroidals companion to the Milky Way (taken from the literature compilation by \citealt{kalirai09}). There does not seem to be any major displacement of our target dwarfs from a linear relation that represents Local Group dwarfs. This relation will however be considered again with the addition of data for late--type dwarfs of the Centaurus A group in two forthcoming papers (Crnojevi\'{c}, Grebel \& Cole, in prep.).

The other panel in Fig. \ref{fehvs} represents the dependence of metallicity on the deprojected distance from Centaurus A (black circles), and the same is reported also for Milky Way dwarf spheroidals (red circles). We compute the distances from the host galaxy by considering the radial distances listed in \citet{kara07} and the angular ones derived by \citet{kara02}. In this relation, our dwarf spheroidals seem to resamble what is seen for Milky Way companions: neither the Centaurus A nor the Milky Way companions show any obvious trend with galactocentric distance from their massive primary (other than increased scatter in close proximity to a massive galaxy). The two dwarf ellipticals KK197 and ESO269-66 are slightly more metal--rich, similarly to the dwarf elliptical companions of M31 (there are no such objects in the vicinity of the Milky Way). Finally, we also look for a possible correlation of the metallicity with the tidal index (i.e., a measure of the degree of isolation of a galaxy, with higher values corresponding to denser environments, \citealt{kara04}), but find no evidence for any trend.

The last aspect investigated in our study is the possible presence of subpopulations with different metallicity content. For the Local Group, photometric, detailed spectroscopic, and in some cases also kinematic studies revealed dwarfs with distinct subpopulations, but also dwarfs that do not present any such trend, and only weak metallicity gradients. It it thus not surprising to find such a diversity also in the Centaurus A group. Recent simulations by \citet{marcolini08} provide a convincing model for the formation of population gradients and can even account for the formation of subpopulations distinct in metallicity and kinematics. According to this model, population gradients are the natural consequence of ``the chemical homogeinization of the interstellar matter, together with the combined inhomogeneous pollution by supernovae type Ia'', provided that star formation lasts for several Gyr. In our sample we have two cases (KK197 and ESO269-66) where the metallicities are quite high, and where moderate metallicity gradients are found. Defining a metal--poor and a metal--rich subsample of stars, we do find evidence for at least two subpopulations which have different radial distributions, even though our separation between the subsamples is arbitrary. This could be the effect of a low, extended star formation at the early stages of these galaxies' life, which would enrich the subsequent stellar populations. Since these are the two most luminous galaxies, the retention of some neutral gas due to deeper potential wells could in principle help keeping the star formation active for some Gyr. The possible existence of small age gradients in old populations with ages $>10$ Gyr cannot be resolved in our data, since the effect of age on the isochrones is too small and we have no other way to disentangle age and metallicity on the upper RGB, but it is almost certainly present and could explain the observed metallicity spreads and gradients.


\section{Conclusions} \label{conclus}

We have analyzed stellar point source photometry for six early--type dwarf galaxies in the Centaurus A group (at an average distance of $\sim3.8$ Mpc), using archival HST/ACS data. Their color--magnitude diagrams show broad red giant branches without any evidence of young ($<1$ Gyr) populations, so we assume a predominantly old age for their stellar content. We note, however, that our target galaxies also exhibit a small number of luminous asymptotic giant branch stars above the tip of their red giant branches, indicative of some contamination with intermediate--age stars (in the range $\sim4-8$ Gyr). In order to derive photometric metallicities for the luminous red giants in our sample, we choose Dartmouth isochrones \citep{dotter08}, since they have been shown to provide an excellent fit to the full extent of color--magnitude diagrams of both old and intermediate--age star clusters. Fixing the age of the isochrones at 10 Gyr and assuming a solar scaled $\alpha$-element abundance, we let the metallicity vary and interpolate between the isochrones to get individual metallicity values for stars on the red giant branch. The results show in all the cases: 1. An on average \emph{metal--poor stellar content} ($<$[Fe/H]$> =-1.56$ to $-1.08$), and 2. a \emph{spread of metallicities} (internal dispersion of $0.10$--$0.42$ dex). We further estimate the amount of bias in our results due to the presence of intermediate--age stars (which account for $\sim10$ to $\sim20\%$ of the entire populations), and find that for two of the target galaxies the metallicity spreads are likely to be the result of this contamination. For the other objects, we show that the presence of extended early star formation episodes (which are indeed very likely) and of a range of $\alpha$-element abundances would eventually broaden the derived metallicity spreads, and make the median metallicities lower by $\sim15\%$.

We investigate the possible presence of metallicity gradients as a function of elliptical galactic radius. We find a moderate overall gradient for one dwarf elliptical ($-0.17$ dex per arcmin, or $-0.15$ dex per kpc), and hints for a gradient in the inner regions of two dwarf spheroidals. We further investigate possible differences in the stellar spatial distributions, splitting the stellar content of each galaxy into metal--poor and metal--rich stars. For the two dwarf ellipticals, which are also the most luminous galaxies of our sample, we can reject the null hypotesis that the two subpopulations come from the same parent distribution with a high statistical confidence level, with the more metal--rich stars begin more centrally concentrated. 

A detailed comparison to well-studied dwarfs in the Local Group is not possible due to the intrinsic uncertainties of our methods. However, the results derived in this study (metal--poor stellar contents, broad metallicity spreads, metallicity gradients) are overall very similar to what is found for the early--type dwarfs in our Local Group, despite the fact that the Centaurus A group is a denser environment, possibly in a more evolved dynamical stage. Having a look at the global properties of the sample, we find that our preliminary metallicity--luminosity relation is similar to the one that holds in the Local Group. We do not find any clear trend of median galaxy metallicity versus tidal index (i.e. degree of isolation), or versus distance from the central galaxy of the group. 

As a next step of this study, in two companion papers we will consider the dwarf irregular galaxies found in the Centaurus A group, and for which HST/ACS data are available. Studying dwarf galaxies in Centaurus A and other groups, and comparing the results to what we already know about the Local Group, will ultimately help us to put further constraints on the connection between star formation, chemical enrichment and environmental effects on the evolution of dwarf galaxies.


\begin{acknowledgements}

DC is very thankful to S. Pasetto, K. Glatt, K. Jordi, S. Jin and T. Lisker for enlightening discussions and useful suggestions. DC acknowledges financial support from the MPIA of Heidelberg, as part of the IMPRS program. AK acknowledges support by an STFC Postdoctoral Fellowship and by the Heidelberg Graduate School for Fundamental Physics of the University of Heidelberg (grant number GSC 129/1). This work is based on observations made with the NASA/ESA Hubble Space Telescope, obtained from the data archive at the Space Telescope Science Institute. STScI is operated by the Association of Universities for Research in Astronomy, Inc. under NASA contract NAS 5-26555.

\end{acknowledgements}


\bibliographystyle{aa}
\bibliography{biblio.bib}



\end{document}